\documentclass[amsmath,aps,showpacs,a4paper,10pt]{revtex4}

 \usepackage{epsf}
 \usepackage{graphicx}    

\begin{document}

 \newcommand{\be}[1]{\begin{equation}\label{#1}}
 \newcommand{\ee}{\end{equation}}
 \newcommand{\bea}{\begin{eqnarray}}
 \newcommand{\eea}{\end{eqnarray}}
 \def\disp{\displaystyle}

 \begin{titlepage}

 \begin{flushright}
 arXiv:2510.09463
 \end{flushright}

 \title{\Large \bf Generalized Distributions of Host Dispersion Measures\\
 in the Fast Radio Burst Cosmology}

 \author{Jing-Yi~Jia\,$^{a,}$\footnote{email
 address:\ jjy@bit.edu.cn}\,,
 Da-Chun~Qiang\,$^b$\,,
 Lin-Yu~Li\,$^a$\,,
 Hao~Wei\,$^{a,}$\footnote{Corresponding author;\ email
 address:\ haowei@bit.edu.cn}\vspace{2.4mm}}
 \affiliation{$^{a)\,}$School of Physics, Beijing
 Institute of Technology, Beijing 100081, China\vspace{2mm}\\
 $^{b)\,}$Institute for Gravitational Wave Astronomy, Henan Academy of
 Sciences, Zhengzhou 450046, Henan, China}

 \begin{abstract}\vspace{1cm}
 \centerline{\bf ABSTRACT}\vspace{2mm}
 Fast radio bursts (FRBs) can provide a measure of the Hubble constant
 $H_0$ that is independent of the constraints set by the cosmic
 microwave background (CMB) and the type Ia supernovae (SNIa), thereby
 arbitrating the Hubble tension.~In the literature, the methodology
 proposed by Macquart {\it et al.} has been widely used, in which the
 contributions to the dispersion measure (DM) from the intergalactic
 medium (IGM, $\rm DM_{IGM}$) and the host galaxy ($\rm DM_{host}$) are
 described by probability distribution functions.~Within the Macquart
 {\it et al.} methodology, it has been found that the parameter $F$,
 which quantifies the strength of the baryon feedback in galaxies, must
 be bound by an artificially narrow prior to result in a Hubble constant
 $H_0$ that is consistent with the ones derived from the CMB and SNIa
 studies.~A recent study using ${\cal O}(100)$ localized FRBs found that
 this also causes the fraction of baryon mass in the IGM, $f_{\rm IGM}$,
 to approach its upper bound of $1$. In the present work, using 125
 localized FRBs, we find an unusually low $H_0$ when using a model with
 a loose prior on $F$.~This model is in fact strongly preferred to the
 model with the narrow prior when considering the Bayesian evidence and
 the Akaike and Bayesian information criteria.~Instead of modifying
 $\sigma_\Delta=Fz^{-0.5}$ in the distribution of $\rm DM_{IGM}$, we
 explore an alternative method of resolving the tension by generalizing
 the distribution of $\rm DM_{host}$ with varying location and scale
 parameters $\ell$ and $e^\mu$, respectively.~We find that $H_0$ can be
 well consistent with the ones of Planck 2018 and SH0ES for all the
 models considered in this work, while these generalized models are
 all strongly preferred to the model with a narrow prior on $F$.~Our
 findings indicate that more realistic distributions of $\rm DM_{host}$
 could be the key to using FRBs as an independent measure of $H_0$.
 \end{abstract}


 \maketitle

 \end{titlepage}

 \renewcommand{\baselinestretch}{1.0}


\section{Introduction}\label{sec1}

The Hubble tension is one of the most serious challenges in cosmology
 to date~\cite{DiValentino:2025sru,Abdalla:2022yfr,Rong-Gen:2023dcz}.~In
 particular, the Hubble constant inferred from the final Planck
 measurements (Planck 2018) of the cosmic microwave background (CMB) in
 the early universe is given by $H_0=67.36\pm 0.54\;{\rm
 km/s/Mpc}$~\cite{Planck:2018vyg}.~On the other hand, the local
 determination of $H_0$ based on the Cepheid/Type Ia supernova~(SNIa)
 distance ladder from the Hubble Space Telescope (HST) and the SH0ES team is
 given by $H_0=73.04\pm 1.04\;{\rm km/s/Mpc}$~\cite{Riess:2021jrx}.~There is
 a serious tension (beyond $5\sigma$) between them.~Many efforts have
 been made in the literature, but this significant discrepancy
 has not been well reconciled yet~\cite{DiValentino:2025sru,Abdalla:2022yfr,
 Rong-Gen:2023dcz}.

It is of interest to measure the Hubble constant by using some
 new probes independent of the CMB and SNIa constraints.~One of the
 promising new probes is fast radio bursts (FRBs)~\cite{NAFRBs,
 Lorimer:2018rwi,Keane:2018jqo,Petroff:2021wug,Xiao:2021omr,
 Zhang:2020qgp,Zhang:2022uzl,Nicastro:2021cxs}, which are transient
 radio sources of millisecond duration whose physical origins are still
 unknown, although might be linked to magnetars, isolated/interacting
 neutron stars, black holes, mergers of compact stars, and so on.~Since
 almost all of FRBs are at extragalactic/cosmological distances (the
 inferred redshifts of FRBs could be as large as $z\sim 3$
 or even larger), they are useful to study cosmology and
 the intergalactic medium (IGM).

One of the key observational quantities of FRBs is the
 dispersion measure (DM), which measures the column density of
 the free electrons, due to the ionized medium (plasma) along
 the path.~The observed DM of an FRB at redshift $z$ can be
 separated into~\cite{Deng:2013aga,Yang:2016zbm,Gao:2014iva,
 Zhou:2014yta,Qiang:2019zrs,Qiang:2020vta,Qiang:2021bwb,Qiang:2021ljr,
 Guo:2022wpf,Guo:2023hgb,Li:2024dge,
 Qiang:2024erm,Qiang:2024lhu}
 \be{eq1}
 {\rm DM_{obs}=DM_{MW,\,ISM}+DM_{MW,\,halo}+DM_{IGM}+DM_{host}}/(1+z)\,,
 \ee
 where $\rm DM_{MW,\,ISM}$, $\rm DM_{MW,\,halo}$, $\rm DM_{IGM}$, and
 $\rm DM_{host}$ are the contributions from the interstellar medium
 (ISM) and halo of our Milky Way (MW), IGM, and the host galaxy
 (including the ISM of the host galaxy and the near-source plasma),
 respectively.~\hspace{0.1em}In particular, $\rm DM_{IGM}$ records the
 main information about the IGM and the cosmic expansion.~The
 mean of $\rm DM_{IGM}$ at redshift $z$ is given
 by~\cite{Deng:2013aga,Yang:2016zbm,Gao:2014iva,Zhou:2014yta,Qiang:2019zrs,
 Qiang:2020vta,Qiang:2021bwb,Qiang:2021ljr,Guo:2022wpf,Guo:2023hgb}
 \be{eq2}
 \langle{\rm DM_{IGM}}\rangle=\frac{3cH_0\Omega_b}{8\pi G m_p}
 \int_0^z\frac{f_{\rm IGM}(\tilde{z})\,f_e(\tilde{z})\left(1+
 \tilde{z}\right)d\tilde{z}}{E(\tilde{z})}\,,
 \ee
 where $c$ is the speed of light, $H_0$ is the Hubble constant,
 $\Omega_b$ is the present fractional density of baryons, $G$
 is the gravitational constant, $m_p$ is the mass of proton,
 $E(z)\equiv H(z)/H_0$ is the dimensionless Hubble parameter,
 $f_{\rm IGM}(z)$ is the fraction of baryon mass in the IGM, and $f_e(z)$ is
 the ionized electron number fraction per baryon.~In principle, one can
 constrain the cosmological parameters, especially the Hubble constant
 $H_0$, by using the observational data of FRBs.

In FRB cosmology, $\rm DM$ plays the role of a proxy of the
 luminosity distance $d_L$.~To study cosmology, the redshift of FRB
 should be known, but this has historically been difficult to determine
 because of the poor FRB localizations and the lack of electromagnetic
 counterparts.~In the absence of measured redshifts, studies originally
 used simulated FRBs with mock redshifts (see e.g.~\cite{Deng:2013aga,
 Yang:2016zbm,Gao:2014iva,Zhou:2014yta,Qiang:2019zrs,Qiang:2020vta,
 Qiang:2021bwb,Qiang:2021ljr,Guo:2022wpf,Guo:2023hgb}).~It is worth
 noting that the method of dark sirens in the field of gravitational
 waves was proposed to be applied to unlocalized
 FRBs in~\cite{Zhao:2022yiv}.~In recent years, more and more
 FRBs have been well localized and hence their redshifts
 could be measured, enabling precision cosmology with FRBs.

Previously, $\rm DM_{host}$ was usually assumed to be a given
 constant, and $\langle{\rm DM_{IGM}}\rangle$ was directly used in place
 of $\rm DM_{IGM}$.~However, these assumptions are not realistic.~Noting
 that $\langle{\rm DM_{IGM}}\rangle$ is the mean value of $\rm DM_{IGM}$
 in all directions of the lines of sight, and Eq.~(\ref{eq2}) is derived
 under the assumption of the cosmological principle, $\rm DM_{IGM}$
 should deviate from $\langle{\rm DM_{IGM}}\rangle$ since the plasma density
 fluctuates along the line of sight~\cite{McQuinn:2013tmc,Ioka:2003fr,
 Inoue:2003ga,Jaroszynski:2018vgh}.~Recently, the new methodology proposed
 by Macquart {\it et al.}~\cite{Macquart:2020lln} has become widely used
 in the literature.~In this methodology, both $\rm DM_{IGM}$ and $\rm
 DM_{host}$ are instead described by probability distribution functions
 (PDFs), and the model parameters are constrained by maximizing the
 likelihood.~We will briefly review this methodology in
 Sec.~\ref{sec2a}.

The Macquart {\it et al.} methodology has worked well in cases in which
 there are few localized FRBs.~For instance, $\Omega_b h_{70}$
 was constrained to $0.051^{+0.021}_{-0.025}$
 ($95\%$ confidence) in~\cite{Macquart:2020lln} by using only 8
 localized FRBs, where $h_{70}=H_0/(70\;{\rm
 km/s/Mpc})$.~\hspace{0.3mm}In e.g.~\cite{Wu:2021jyk}, $H_0=
 68.81^{+4.99}_{-4.33}\; {\rm km/s/Mpc}$ was found by using 18 localized
 FRBs, which is consistent with both the SH0ES and Planck 2018 results.~When
 the number of localized FRBs increased, the situation changed slowly as
 follows.~\hspace{0.3mm}In e.g.~\cite{Kalita:2024xae}, $H_0\simeq 73\sim
 76\;{\rm km/s/Mpc}$ was found by using 64 localized FRBs, which is
 consistent with the SH0ES result but inconsistent with the one
 of Planck 2018.~In e.g.~\cite{Gao:2025fcr}, in order to get a
 Hubble constant $H_0=69.40^{+2.14}_{-1.97}\;{\rm km/s/Mpc}$
 consistent with the Planck 2018 result by using 108 localized FRBs,
 \hspace{-0.2mm}$f_{\rm IGM}=0.93$ had to be adopted, which is much
 larger than \hspace{-0.3mm}$f_{\rm IGM}=0.82\sim 0.84$ extensively used
 in the literature (e.g.~\cite{Deng:2013aga,Yang:2016zbm,
 Gao:2014iva,Zhou:2014yta,Qiang:2019zrs,Qiang:2020vta,Qiang:2021bwb,
 Qiang:2021ljr,Guo:2022wpf,Guo:2023hgb}) and obtained independently from
 the observations of the Ly$\alpha$ forest and UV
 absorption lines~\cite{Fukugita:1997bi,Shull:2011aa}.~In
 e.g.~\cite{Lemos:2025bgy}, adopting the SH0ES result $H_0=73.04\pm
 1.04\;{\rm km/s/Mpc}$, unusually high values of $f_{\rm IGM}=0.935$ to
 $0.999$ were found by using 107 localized FRBs in a more
 cosmology-independent way, which is extremely close to its upper bound
 of $1$.~To date, the debate on $H_0$ and $f_{\rm IGM}$
 from localized FRBs is still not settled.

Another point of tension arises from the constraints on
 $\sigma_\Delta=Fz^{-0.5}$ in the Macquart {\it et al.} methodology
 which characterizes the galactic feedback (see Sec.~\ref{sec2a} for
 details).~A narrow prior $[\,0.011,\,0.5\,]$ for the parameter $F$ was
 used by Macquart {\it et al.}~\cite{Macquart:2020lln}, where the upper
 bound of $F$ is only $0.5$.~With this choice of prior, $F$ cannot be
 constrained from the right hand side (see Extended Data Fig.~5
 of~\cite{Macquart:2020lln}).~Only 8 localized FRBs were used in this study,
 however, the issue with the constraint on $F$ still holds with a larger
 data set.~As shown in e.g.~Fig.~5 of~\cite{Zhuge:2025urk}, $F$ still cannot
 be constrained from the right hand side even when using 115 localized
 FRBs.~In fact, if the prior for $F$ could be relaxed to
 e.g.~$[\,0.01,\,10.0\,]$, $F$ will be well constrained, but a much
 larger value of $F\sim {\cal O}(1)$ is favored (see Sec.~\ref{sec2c}
 below), and the corresponding constraints on the model
 parameters (especially $H_0$) will be significantly changed.~When exploring
 the effect of $F$ on $H_0$, as discussed in
 e.g.~\cite{Baptista:2023uqu} and Sec.~4.4 of~\cite{Xu:2025ddk}, it
 was found that a small $F$ is required to obtain a Hubble constant
 $H_0$ consistent with the ones of Planck 2018 and SH0ES.~Noting that
 $\sigma_\Delta=Fz^{-0.5}$ is related to the variance of the
 distribution of $\rm DM_{IGM}$, one of the possible ways out is to
 modify $\sigma_\Delta$ as in e.g.~\cite{Zhuge:2025urk}.

In the present work, we test the robustness
 of the Macquart~{\it et~al.}~methodology~\cite{Macquart:2020lln}, by
 allowing for more general distributions of $\rm DM_{host}$,
 while simultaneously addressing its limitation (e.g.~$F$ is unbounded
 from above) and also alleviating the tension between the constraints on
 $H_0$ determined using FRBs with the constraints from the CMB and SNIa
 measurements.~In Sec.~\ref{sec2}, we briefly review the Macquart {\it
 et al.} methodology.~\hspace{0.36mm}In Sec.~\ref{sec2c}, we apply the
 Macquart {\it et al.} methodology to the fiducial model with a loose
 $F$ prior $[\,0.01,\,10.0\,]$ and the NarrowF model with
 the same priors used by Macquart {\it et al.}~\cite{Macquart:2020lln}
 (especially the narrow $F$ prior $[\,0.011,\,0.5\,]$).~We argue that
 the Hubble tension between the CMB and SNIa constraints could
 be resolved with FRB measurements, but the assumptions must
 be considered more carefully to produce reliable results.~\hspace{0.36mm}In
 Secs.~\ref{sec3} and \ref{sec4}, we consider the generalized
 distributions of $\rm DM_{host}$ with varying location and scale parameters
 $\ell$ and $e^\mu$, respectively.~We will check whether the
 great Hubble tension between FRBs, Planck 2018 and SH0ES could
 be alleviated in these generalized models.~Finally, some brief
 concluding remarks are given in Sec.~\ref{sec5}.


 \begingroup
 \renewcommand{\baselinestretch}{0.5}
 \squeezetable
 \begin{table}[pth]
 \begin{center}
 \vspace{-14.1mm}  
 \hspace{-7mm}  
 \begin{tabular}{lcccccc|clccccc} \hline\hline
  & & & & & & & \hspace{0.3mm} & & & & & & \\[-2.7mm]
 ~~~~FRB & R.A. & Dec. & $\rm DM_{obs}$ & Redshift & Ref. & & & ~~~~FRB & R.A. & Dec. & $\rm DM_{obs}$ & Redshift & Ref.\\[1mm] \hline
  & & & & & & & & & & & & & \\[-2.8mm]
 20220207C~ & 310.1995 & 72.8823 & 262.38 & 0.04304 & \cite{Law:2023ibd,Sharma:2024fsq} & & & 20220307B~ & 350.8745 & 72.1924 & 499.27 & 0.248123 & \cite{Law:2023ibd,Sharma:2024fsq} \\
 20220310F & 134.7204 & 73.4908 & 462.24 & 0.477958 & \cite{Law:2023ibd,Sharma:2024fsq} & & & 20220319D & 32.1779 & 71.0353 & 110.98 & 0.011228 & \cite{Law:2023ibd,Sharma:2024fsq} \\
 20220418A & 219.1056 & 70.0959 & 623.25 & 0.622 & \cite{Law:2023ibd,Sharma:2024fsq} & & & 20220506D & 318.0448 & 72.8273 & 396.97 & 0.30039 & \cite{Law:2023ibd,Sharma:2024fsq} \\
 20220509G & 282.67 & 70.2438 & 269.53 & 0.0894 & \cite{Law:2023ibd,Sharma:2024fsq} & & & 20220825A & 311.9815 & 72.5850 & 651.24 & 0.241397 & \cite{Law:2023ibd,Sharma:2024fsq} \\
 20220914A & 282.0568 & 73.3369 & 631.28 & 0.1139 & \cite{Law:2023ibd,Sharma:2024fsq} & & & 20220920A & 240.2571 & 70.9188 & 314.99 & 0.158239 & \cite{Law:2023ibd,Sharma:2024fsq} \\
 20221012A & 280.7987 & 70.5242 & 441.08 & 0.284669 & \cite{Law:2023ibd,Sharma:2024fsq} & & & 20220912A & 347.2704 & 48.7071 & 219.46 & 0.0771 & \cite{DeepSynopticArrayTeam:2022rbq,Zhang:2023eui} \\
 20210117A & 339.9792 & $-16.1515$ & 729.1 & 0.214 & \cite{Bhandari:2022ton} & & & 20181220A & 348.6982 & 48.3421 & 208.66 & 0.02746 & \cite{Bhardwaj:2023vha,CHIMEFRB:2021srp} \\
 20181223C & 180.9207 & 27.5476 & 111.61 & 0.03024 & \cite{Bhardwaj:2023vha,CHIMEFRB:2021srp} & & & 20190418A & 65.8123 & 16.0738 & 182.78 & 0.07132 & \cite{Bhardwaj:2023vha,CHIMEFRB:2021srp} \\
 20190425A & 255.6625 & 21.5767 & 127.78 & 0.03122 & \cite{Bhardwaj:2023vha,CHIMEFRB:2021srp} & & & 20220610A & 351.0732 & $-33.5137$ & 1458.15 & 1.016 & \cite{Ryder:2022qpg} \\
 20200120E & 149.4863 & 68.8256 & 87.782 & $-0.0001$ & \cite{Bhardwaj:2021xaa} & & & 20171020A & 333.75 & $-19.6667$ & 114.1 & 0.008672 & \cite{Mahony:2018ddp} \\
 20121102A & 82.9946 & 33.1479 & 557.0 & 0.1927 & \cite{Gordon:2023cgw} & & & 20180301A & 93.2268 & 4.6711 & 536.0 & 0.3304 & \cite{Gordon:2023cgw,Price:2019fmc} \\
 20180916B & 29.5031 & 65.7168 & 347.8 & 0.0337 & \cite{Gordon:2023cgw,CHIMEFRB:2021srp} & & & 20180924B & 326.1053 & $-40.9000$ & 362.42 & 0.3212 & \cite{Gordon:2023cgw} \\
 20181112A & 327.3485 & $-52.9709$ & 589.27 & 0.4755 & \cite{Gordon:2023cgw} & & & 20190102C & 322.4157 & $-79.4757$ & 363.6 & 0.2912 & \cite{Gordon:2023cgw} \\
 20190608B & 334.0199 & $-7.8982$ & 338.7 & 0.1178 & \cite{Hiramatsu:2022tyn,Gordon:2023cgw} & & & 20190611B & 320.7456 & $-79.3976$ & 321.4 & 0.3778 & \cite{Gordon:2023cgw} \\
 20190711A & 329.4192 & $-80.358$ & 593.1 & 0.522 & \cite{Gordon:2023cgw,Macquart:2020lln} & & & 20190714A & 183.9797 & $-13.021$ & 504.13 & 0.2365 & \cite{Hiramatsu:2022tyn,Gordon:2023cgw,HESS:2021smp,Guidorzi:2020ggq} \\
 20191001A & 323.3513 & $-54.7478$ & 506.92 & 0.234 & \cite{Gordon:2023cgw,Bhandari:2020cde} & & & 20200430A & 229.7064 & 12.3767 & 380.1 & 0.1608 & \cite{Hiramatsu:2022tyn,Gordon:2023cgw} \\
 20200906A & 53.4962 & $-14.0832$ & 577.8 & 0.3688 & \cite{Hiramatsu:2022tyn,Gordon:2023cgw} & & & 20201124A & 77.0146 & 26.0607 & 415.3 & 0.0979 & \cite{Lanman:2021yba,Gordon:2023cgw} \\
 20210320C & 204.4608 & $-16.1227$ & 384.8 & 0.2797 & \cite{Gordon:2023cgw} & & & 20210410D & 326.0863 & $-79.3182$ & 571.2 & 0.1415 & \cite{Caleb:2023atr,Gordon:2023cgw} \\
 20210807D & 299.2214 & $-0.7624$ & 251.9 & 0.1293 & \cite{Gordon:2023cgw} & & & 20211127I & 199.8082 & $-18.8378$ & 234.83 & 0.0469 & \cite{Gordon:2023cgw} \\
 20211203C & 204.5625 & $-31.3801$ & 636.2 & 0.3439 & \cite{Gordon:2023cgw} & & & 20211212A & 157.3509 & 1.3609 & 206.0 & 0.0707 & \cite{Gordon:2023cgw} \\
 20220105A & 208.8039 & 22.4665 & 583.0 & 0.2785 & \cite{Gordon:2023cgw} & & & 20191106C & 199.5801 & 42.9997 & 332.2 & 0.10775 & \cite{Ibik:2023ugl,CHIMEFRB:2021srp} \\
 20200223B & 8.2695 & 28.8313 & 201.8 & 0.06024 & \cite{Ibik:2023ugl,CHIMEFRB:2021srp} & & & 20190110C & 249.3185 & 41.4434 & 221.6 & 0.12244 & \cite{Ibik:2023ugl,CHIMEFRB:2021srp} \\
 20190303A & 207.9958 & 48.1211 & 223.2 & 0.064 & \cite{Michilli:2022bbs,CHIMEFRB:2021srp} & & & 20180814A & 65.6833 & 73.6644 & 190.9 & 0.068 & \cite{Michilli:2022bbs,CHIMEFRB:2021srp} \\
 20210405I & 255.3397 & $-49.5451$ & 565.17 & 0.066 & \cite{Driessen:2023lxj} & & & 20191228A & 344.4304 & $-28.5941$ & 297.5 & 0.2432 & \cite{Bhandari:2021pvj} \\
 20181030A & 158.5838 & 73.7514 & 103.5 & 0.00385 & \cite{Bhardwaj:2021hgc,CHIMEFRB:2021srp} & & & 20190523A & 207.065 & 72.4697 & 760.8 & 0.66 & \cite{Ravi:2019alc} \\
 20190614D & 65.0755 & 73.7067 & 959.2 & 0.6 & \cite{Law:2020cnm,Hiramatsu:2022tyn} & & & 20210603A & 10.2741 & 21.2263 & 500.147 & 0.1772 & \cite{Cassanelli:2023hvg} \\
 20231120A & 143.9840 & 73.2847 & 437.737 & 0.0368 & \cite{Sharma:2024fsq} & & & 20230124A & 231.9162 & 70.9681 & 590.574 & 0.0939 & \cite{Sharma:2024fsq} \\
 20230628A & 166.7867 & 72.2818 & 344.952 & 0.127 & \cite{Sharma:2024fsq} & & & 20221101B & 342.2162 & 70.6812 & 491.554 & 0.2395 & \cite{Sharma:2024fsq} \\
 20221113A & 71.411 & 70.3074 & 411.027 & 0.2505 & \cite{Sharma:2024fsq} & & & 20231123B & 242.5382 & 70.7851 & 396.857 & 0.2621 & \cite{Sharma:2024fsq} \\
 20230307A & 177.7813 & 71.6956 & 608.854 & 0.2706 & \cite{Sharma:2024fsq} & & & 20221116A & 21.2102 & 72.6539 & 643.448 & 0.2764 & \cite{Sharma:2024fsq} \\
 20230501A & 340.0272 & 70.9222 & 532.471 & 0.3015 & \cite{Sharma:2024fsq} & & & 20230626A & 235.6296 & 71.1335 & 452.723 & 0.327 & \cite{Sharma:2024fsq} \\
 20220208A & 322.5751 & 70.0410 & 440.73 & 0.351 & \cite{Sharma:2024fsq} & & & 20220726A & 73.9457 & 69.9291 & 686.232 & 0.3619 & \cite{Sharma:2024fsq} \\
 20220330D & 163.7512 & 70.3508 & 467.788 & 0.3714 & \cite{Sharma:2024fsq} & & & 20220204A & 274.2262 & 69.7225 & 612.584 & 0.4012 & \cite{Sharma:2024fsq} \\
 20230712A & 167.3585 & 72.5578 & 587.567 & 0.4525 & \cite{Sharma:2024fsq} & & & 20230216A & 156.4722 & 3.4368 & 828.289 & 0.531 & \cite{Sharma:2024fsq} \\
 20221027A & 130.8718 & 72.1010 & 452.723 & 0.5422 & \cite{Sharma:2024fsq} & & & 20221219A & 257.6298 & 71.6268 & 706.708 & 0.553 & \cite{Sharma:2024fsq} \\
 20221029A & 141.9634 & 72.4523 & 1391.746 & 0.975 & \cite{Sharma:2024fsq} & & & 20240114A & 321.9161 & 4.3292 & 527.65 & 0.13 & \cite{Tian:2024ygd} \\
 20220501C & 352.3792 & $-32.4907$ & 449.5 & 0.381 & \cite{Shannon:2024pbu} & & & 20220725A & 353.3152 & $-35.9903$ & 290.4 & 0.1926 & \cite{Shannon:2024pbu} \\
 20220918A & 17.5921 & $-70.8114$ & 656.8 & 0.491 & \cite{Shannon:2024pbu} & & & 20221106A & 56.7048 & $-25.5698$ & 343.8 & 0.2044 & \cite{Shannon:2024pbu} \\
 20230526A & 22.2326 & $-52.7173$ & 361.4 & 0.157 & \cite{Shannon:2024pbu} & & & 20230708A & 303.1155 & $-55.3563$ & 411.51 & 0.105 & \cite{Shannon:2024pbu} \\
 20230902A & 52.1398 & $-47.3335$ & 440.1 & 0.3619 & \cite{Shannon:2024pbu} & & & 20231226A & 155.3638 & 6.1102 & 329.9 & 0.1569 & \cite{Shannon:2024pbu} \\
 20240201A & 149.9056 & 14.0880 & 374.5 & 0.042729 & \cite{Shannon:2024pbu} & & & 20240208A & 159.2296 & $-0.9544$ & 260.2 & 0.39 & \cite{Shannon:2024pbu} \\
 20240210A & 8.7796 & $-28.2707$ & 283.73 & 0.023686 & \cite{Shannon:2024pbu} & & & 20240310A & 17.6219 & $-44.4394$ & 601.8 & 0.127 & \cite{Shannon:2024pbu} \\
 20240318A & 150.3932 & 37.6164 & 256.4 & 0.12 & \cite{Shannon:2024pbu} & & & 20230718A & 128.1619 & $-40.4519$ & 477.0 & 0.035 & \cite{Shannon:2024pbu,Arcus:2024pvz,Glowacki:2024} \\
 20201123A & 263.67 & $-50.76$ & 433.55 & 0.0507 & \cite{Rajwade:2022zkj,Kalita:2024kyo} & & & 20230521B & 351.036 & 71.1380 & 1342.9 & 1.354 & \cite{Connor:2024mjg} \\
 20230814B & 335.9747 & 73.0259 & 696.4 & 0.553 & \cite{Connor:2024mjg} & & & 20231220A & 123.9087 & 73.6599 & 491.2 & 0.3355 & \cite{Connor:2024mjg} \\
 20240119A & 224.4672 & 71.6118 & 483.1 & 0.376 & \cite{Connor:2024mjg} & & & 20240123A & 68.2625 & 71.9453 & 1462.0 & 0.968 & \cite{Connor:2024mjg} \\
 20240213A & 166.1683 & 74.0754 & 357.4 & 0.1185 & \cite{Connor:2024mjg} & & & 20240215A & 268.4413 & 70.2324 & 549.5 & 0.21 & \cite{Connor:2024mjg} \\
 20240229A & 169.9835 & 70.6762 & 491.15 & 0.287 & \cite{Connor:2024mjg} & & & 20230203A & 151.6616 & 35.6941 & 420.1 & 0.1464 & \cite{CHIMEFRB:2025ggb} \\
 20230222A & 106.9604 & 11.2245 & 706.1 & 0.1223 & \cite{CHIMEFRB:2025ggb} & & & 20230222B & 238.7391 & 30.8987 & 187.8 & 0.11 & \cite{CHIMEFRB:2025ggb} \\
 20230311A & 91.1097 & 55.9460 & 364.3 & 0.1918 & \cite{CHIMEFRB:2025ggb} & & & 20230703A & 184.6244 & 48.7299 & 291.3 & 0.1184 & \cite{CHIMEFRB:2025ggb} \\
 20230730A & 54.6646 & 33.1593 & 312.5 & 0.2115 & \cite{CHIMEFRB:2025ggb} & & & 20230926A & 269.1249 & 41.8143 & 222.8 & 0.0553 & \cite{CHIMEFRB:2025ggb} \\
 20231005A & 246.028 & 35.4487 & 189.4 & 0.0713 & \cite{CHIMEFRB:2025ggb} & & & 20231011A & 18.2411 & 41.7491 & 186.3 & 0.0783 & \cite{CHIMEFRB:2025ggb} \\
 20231017A & 346.7543 & 36.6527 & 344.2 & 0.245 & \cite{CHIMEFRB:2025ggb} & & & 20231025B & 270.7881 & 63.9891 & 368.7 & 0.3238 & \cite{CHIMEFRB:2025ggb} \\
 20231123A & 82.6232 & 4.4755 & 302.1 & 0.0729 & \cite{CHIMEFRB:2025ggb} & & & 20231128A & 199.5782 & 42.9927 & 331.6 & 0.1079 & \cite{CHIMEFRB:2025ggb} \\
 20231201A & 54.5893 & 26.8177 & 169.4 & 0.1119 & \cite{CHIMEFRB:2025ggb} & & & 20231204A & 207.9990 & 48.116 & 221.0 & 0.0644 & \cite{CHIMEFRB:2025ggb} \\
 20231206A & 112.4428 & 56.2563 & 457.7 & 0.0659 & \cite{CHIMEFRB:2025ggb} & & & 20231223C & 259.5446 & 29.4979 & 165.8 & 0.1059 & \cite{CHIMEFRB:2025ggb} \\
 20231229A & 26.4678 & 35.1129 & 198.5 & 0.019 & \cite{CHIMEFRB:2025ggb} & & & 20231230A & 72.7976 & 2.3940 & 131.4 & 0.0298 & \cite{CHIMEFRB:2025ggb} \\
 20220717A & 293.3042 & $-19.2877$ & 637.34 & 0.36295 & \cite{Wang:2025ugc} & & & 20220529A & 19.1042 & 20.6325 & 246.3 & 0.1839 & \cite{Wang:2025ugc,Li:2025ckl} \\
 20240124A & 321.9162 & 4.3501 & 526.9 & 0.269 & \cite{Piratova-Moreno:2025cpc} & & & 20220222C & 203.9045 & $-28.0269$ & 1071.2 & 0.853 & \cite{Pastor-Marazuela:2025loc} \\
 20220224C & 166.6775 & $-22.9399$ & 1140.2 & 0.6271 & \cite{Pastor-Marazuela:2025loc} & & & 20230125D & 150.2050 & $-31.5447$ & 640.08 & 0.3265 & \cite{Pastor-Marazuela:2025loc} \\
 20230503E & 238.4300 & $-83.7753$ & 483.74 & 0.32 & \cite{Pastor-Marazuela:2025loc} & & & 20230613A & 356.8527 & $-27.0528$ & 483.51 & 0.3923 & \cite{Pastor-Marazuela:2025loc} \\
 20230907D & 187.1425 & 8.6581 & 1030.79 & 0.4638 & \cite{Pastor-Marazuela:2025loc} & & & 20231010A & 14.7320 & $-70.5964$ & 442.59 & 0.61 & \cite{Pastor-Marazuela:2025loc} \\
 20231020B & 57.2782 & $-37.7699$ & 952.2 & 0.4775 & \cite{Pastor-Marazuela:2025loc} & & & 20231210F & 50.4053 & $-35.7614$ & 720.6 & 0.5 & \cite{Pastor-Marazuela:2025loc} \\
 20190520B & 240.5178 & $-11.2881$ & 1201.0 & 0.2414 & \cite{Gordon:2023cgw} & & & 20220831A & 338.6955 & 70.5384 & 1146.25 & 0.262 & \cite{Connor:2024mjg} \\
 20240304B & 182.997 & 11.813 & 2458.2 & 2.148 & \cite{Caleb:2025uzd} & & & 20250316A & 182.4347 & 58.8491 & 161.82 & 0.0067 & \cite{CHIME:2025mlf} \\
 20241228A & 216.3857 & 12.0250 & 246.3 & 0.1614 & \cite{Curtin:2025tvg} & & & \\[1.2mm]
 \hline\hline
 \end{tabular}
 \end{center}
 \vspace{-5mm}  
 \caption{\label{tab1} The sample of current 131 localized
 FRBs.~R.A./Dec.~and $\rm DM_{obs}$ are in units of degree and
 ${\rm pc\hspace{0.24em} cm^{-3}}$, respectively.~See the text for details.}
 \end{table}
 \endgroup



\section{The fiducial methodology}\label{sec2}


\subsection{The methodology and the data}\label{sec2a}

First, we follow the Macquart {\it et al.}
 methodology~\cite{Macquart:2020lln} to constrain the Hubble constant
 $H_0$ by using the currently known localized FRBs.~The starting point
 is Eq.~(\ref{eq1}).~For a given FRB with known right ascension (R.A.)
 and declination (Dec.), its $\rm DM_{MW,\,ISM}$
 and $\rm DM_{MW,\,halo}$ can be found by using NE2001~\cite{Cordes:2002wz,
 Cordes:2003ik} and YT2020~\cite{Yamasaki:2019htx}, respectively.~To
 this end, we use the Python package PyGEDM~\cite{pygedm,Price:2021gzo}
 incorporating NE2001/YMW16 and YT2020.~Note that in this way
 $\rm DM_{MW,\,halo}$ is not a constant, slightly different
 from~\cite{Macquart:2020lln}.~It is convenient to introduce
 the extragalactic DM, namely
 \be{eq3}
 {\rm DM_E\equiv DM_{obs}-DM_{MW,\,ISM}-DM_{MW,\,halo}=
 DM_{IGM}+DM_{host}}/(1+z)\,,
 \ee
 where we used Eq.~(\ref{eq1}) in the second step.~The mean of
 $\rm DM_{IGM}$ at redshift $z$ in Eq.~(\ref{eq2}) can be recast as
 \be{eq4}
 \langle{\rm DM_{IGM}}\rangle=\frac{3cf_e\cdot\Theta}{8\pi G
 m_p}\int_0^z\frac{\left(1+\tilde{z}\right)d\tilde{z}}{E(\tilde{z})}\,,
 \quad\quad
 \Theta\equiv\Omega_b H_0 f_{\rm IGM}=\frac{(\Omega_b h^2)\,
 f_{\rm IGM}}{H_0}\cdot\left(100\;{\rm km/s/Mpc}\right)^2\,,
 \ee
 where $f_e=7/8$ as in e.g.~\cite{Deng:2013aga,Yang:2016zbm,Gao:2014iva,
 Zhou:2014yta,Qiang:2019zrs,Qiang:2020vta,Qiang:2021bwb,Qiang:2021ljr,
 Guo:2022wpf,Guo:2023hgb}, $h=H_0/(100\;{\rm km/s/Mpc})$, $\Theta$ is in
 units of ${\rm km/s/Mpc}$, and
 \be{eq5}
 E(z)\equiv H(z)/H_0=\sqrt{\Omega_m\,(1+z)^3+(1-\Omega_m)}\,,
 \ee
 for a flat $\Lambda$CDM cosmology in the Friedmann-Robertson-Walker
 (FRW) universe.~In this work, we adopt $\Omega_m=0.3153$ from
 the Planck 2018 result~\cite{Planck:2018vyg}.~The combination $\Theta$
 characterizes the degeneracy between $H_0$, $\Omega_b h^2$
 and $f_{\rm IGM}$.~We regard $\Theta$ as a free model parameter, and
 $H_0$ can be derived from $\Theta$ if $\Omega_b h^2$ and $f_{\rm IGM}$
 are given.~As mentioned above, $\rm DM_{IGM}$ fluctuates
 around $\langle{\rm DM_{IGM}}\rangle$ due to the plasma density fluctuation
 along the line of sight.~The distribution of $\rm DM_{IGM}$ is described by
 the PDF~\cite{Macquart:2020lln}
 \be{eq6}
 P_{\rm IGM}(\Delta)=A\Delta^{-\beta}\exp\left[-\frac{\left(\Delta^{-\alpha}
 -C_0\right)^2}{2\alpha^2\sigma^2_\Delta}\right]\quad\quad {\rm for}
 \quad \Delta>0\,,
 \ee
 and $P_{\rm IGM}(\Delta)=0$ for $\Delta\leq 0$,
 where $\Delta\equiv {\rm DM_{IGM}}/\langle{\rm DM_{IGM}}\rangle$, and
 $\alpha=\beta=3$ as in~\cite{Macquart:2020lln} (it was found
 that $\alpha=\beta=3$ provide the best match
 in~\cite{Macquart:2020lln}).~Following~\cite{Macquart:2020lln}, one
 can adopt the parameterization~\cite{McQuinn:2013tmc}


 \begin{table}[tb]
 \renewcommand{\arraystretch}{1.6}
 \begin{center}
 \vspace{-5mm}  
 \begin{tabular}{c} \hline\hline
 $\left|\,\Delta {\rm AIC}\,\right|$\\ \hline
 \renewcommand{\arraystretch}{1.6}
 \quad\ Level of empirical support for the model with the smaller AIC\quad\ \ \\ \hline
 \renewcommand{\arraystretch}{1.0}
 \begin{tabular}{ccc}\\[-3.6mm]
 $0-2$\hspace{10mm} &$4-7$\hspace{10mm} &$>10$\\[-0.3mm]
 Weak\hspace{10mm} &Mild\hspace{10mm} &Strong\\[1mm]
 \end{tabular}
 \\ \hline\hline
 $\left|\,\Delta {\rm BIC}\,\right|$\\ \hline
 \renewcommand{\arraystretch}{1.6}
 Evidence against the model with the larger BIC\\ \hline
 \renewcommand{\arraystretch}{1.0}
 \begin{tabular}{cccc}\\[-3.6mm]
 $0-2$\hspace{10mm} &$2-6$\hspace{10mm} &$6-10$\hspace{10mm} &$>10$\\[-0.3mm]
 Weak\hspace{10mm} &Positive\hspace{10mm} &Strong\hspace{10mm} &Very strong\\[1mm]
 \end{tabular}
 \\ \hline\hline
 $\left|\,\ln {\cal B}\,\right|$\\ \hline
 \renewcommand{\arraystretch}{1.6}
 Evidence against the model with the smaller $\cal Z$\\ \hline
 \renewcommand{\arraystretch}{1.0}
 \begin{tabular}{cccc}\\[-3.6mm]
 $0-1$\hspace{10mm} & $1-2.5$\hspace{10mm} & $2.5-5$\hspace{10mm} & $>5$\\[-0.3mm]
 Inconclusive\hspace{10mm} &Weak\hspace{10mm} & Moderate\hspace{10mm} & Strong\\[1mm]
 \end{tabular}
 \\ \hline\hline
 \end{tabular}
 \end{center}
 \vspace{-1mm}  
 \caption{\label{tab2} The empirical strength of
 $\Delta$AIC, $\Delta$BIC~\cite{Perivolaropoulos:2022khd} and
 $\ln {\cal B}$~\cite{Kilbinger:2009by,Heavens:2017hkr} (see
 also e.g.~\cite{Jia:2025prq}).}
 \end{table}


\vspace{-3mm}  

 \be{eq7}
 \sigma_\Delta=F z^{-0.5}\,,
 \ee
 which characterizes the galactic feedback, and $F$ quantifies
 the strength of the baryon feedback~\cite{McQuinn:2013tmc,
 Macquart:2020lln}. $C_0$ is determined by requiring $\langle\Delta
 \rangle=1$, and we can find $A$ from the normalization of
 $P_{\rm IGM}$.~The rest frame $\rm DM_{host}$~\hspace{-0.5mm}(in units
 of ${\rm pc\hspace{0.24em} cm^{-3}}$) is described by a
 log-normal distribution~\cite{Macquart:2020lln}
 \be{eq8}
 P_{\rm host}({\rm DM_{host}}\,|\,\mu,\,\sigma_{\rm host})=
 \frac{1}{\sigma_{\rm host}{\rm DM_{host}}\sqrt{2\pi}}\cdot\exp
 \left[-\frac{\left(\,\log {\rm DM_{host}}-\mu\right)^2}
 {2\sigma_{\rm host}^2}\,\right]\quad\quad {\rm for}\quad
 {\rm DM_{host}}>0\,,
 \ee
 and $P_{\rm host}({\rm DM_{host}})=0$ for $\rm DM_{host}\leq 0$,~where
 ``\,$\log$\,'' gives the natural logarithm (note that in the rest of
 this work we use ``\,$\ln$\,'' instead).~This distribution has a median
 value of $e^\mu$ (in units of ${\rm pc\hspace{0.24em}
 cm^{-3}}$)~\cite{Macquart:2020lln,lognorm}. Noting Eq.~(\ref{eq3}), for
 a given $i$-th FRB with known redshift $z_i$ and model parameters,
 we have~\cite{Macquart:2020lln,Gao:2025fcr,Zhuge:2025urk}
 \be{eq9}
 P_i({\rm DM}_{{\rm E},\,i}\,|\,z_i)=\int_0^{\,(1+z_i)\,{\rm
 DM}_{{\rm E},\,i}} P_{\rm host}({\rm DM_{host}}\,|\,\mu,\,\sigma_{\rm
 host})\,P_{\rm IGM}({\rm DM}_{{\rm E},\,i}-{\rm DM_{host}}/(1+z_i))\,
 d\,{\rm DM_{host}}\,,
 \ee
 where we have used $0\leq {\rm DM_{host}}=(1+z)({\rm DM_E-DM_{IGM}})
 \leq\left(1+z\right){\rm DM_E}$ from Eq.~(\ref{eq3}), and
 ${\rm DM}_{{\rm E},\,i}={\rm DM}_{{\rm obs},\,i}-{\rm DM}_{{\rm MW,\,
 ISM},\,i}-{\rm DM}_{{\rm MW,\,halo},\,i}\,$.~Note that the
 factor $(1+z)$ has been dropped in~\cite{Macquart:2020lln}, since they
 argued that its effect could be small for their sample of only
 8 localized FRBs.~However, in the accompanying code~\cite{codeM} at
 GitHub for~\cite{Macquart:2020lln}, the factor $(1+z)$ has
 been correctly considered, but in an alternative way (redefining ${\rm
 DM}_{\rm host}^\prime={\rm DM_{host}}/(1+z)$ and then
 recasting Eq.~(\ref{eq9}) in terms of ${\rm
 DM}_{\rm host}^\prime$). Our Eq.~(\ref{eq9}) corresponds to
 e.g.~Eq.~(8) of~\cite{Gao:2025fcr}.~If $y=\eta x$ and $\eta$
 is a constant, $P(y)=(1/\eta)\,P(x)=(1/\eta)\,P(y/\eta)$.~One
 can recast Eq.~(\ref{eq9}) as
 \bea
 &&\hspace{-10mm}P_i({\rm DM}_{{\rm E},\,i}\,|\,z_i) = \nonumber \\
 &&\hspace{-6mm}\frac{1}{\langle{\rm DM_{IGM}}\rangle_i}
 \int_0^{\,(1+z_i)\,{\rm DM}_{{\rm E},\,i}}
 P_{\rm host}({\rm DM_{host}}\,|\,\mu,\,\sigma_{\rm host})\,
 P_{\rm IGM}\left(\frac{{\rm DM}_{{\rm E},\,i}-
 {\rm DM_{host}}/(1+z_i)}{\langle{\rm DM_{IGM}}\rangle_i}\right)d\,
 {\rm DM_{host}}\,,\label{eq10}
 \eea
 where $\langle{\rm DM_{IGM}}\rangle_i$ is $\langle{\rm DM_{IGM}}\rangle$ at
 redshift $z_i$.~Our Eq.~(\ref{eq10}) corresponds
 to e.g.~Eq.~(26) or Eq.~(A7) of~\cite{Zhuge:2025urk} (equivalent to
 Eq.~(A10) of~\cite{Zhuge:2025urk} in terms of $\Delta$).~Note that
 in the accompanying code~\cite{codeM} at GitHub
 for~\cite{Macquart:2020lln}, the pre-factor $1/\langle{\rm DM_{IGM}}
 \rangle$ has been lost, and this point was independently found
 by e.g.~\cite{Zhang:2025wif,Liu:2025fdf} and
 \cite{Zhuge:2025urk}.~We refer to Appendix~A of \cite{Zhuge:2025urk}
 for the detailed derivation.~Finally, the total likelihood is given by
 the joint likelihoods of all localized FRBs~\cite{Macquart:2020lln}, namely
 \be{eq11}
 {\cal L}=\prod_{i=1}^{N_{\rm FRB}} P_i({\rm DM}_{{\rm E},\,i}\,|\,z_i)\,.
 \ee
 We use the Markov Chain Monte Carlo (MCMC) Python package
 Cobaya~\cite{Torrado:2020dgo,Cobaya} with GetDist~\cite{Lewis:2019xzd,
 GetDist} to maximize the likelihood $\cal L$, and then
 obtain the constraints on the model parameters.

In Table~\ref{tab1}, we present the sample of 131 localized
 FRBs.~The redshift of FRB 20200120E is $-0.0001$, due to its peculiar
 velocity towards us, and hence it is decoupled from the cosmic expansion in
 fact.~FRBs 20190520B and 20220831A have extremely large $\rm
 DM_{host}$~\cite{Niu:2021bnl,Connor:2024mjg}, while FRB 20190520B might
 be also influenced by the strong $\rm DM$ from intervening
 galaxies~\cite{Lee:2023}. In addition, we also exclude FRBs 20220319D,
 20210405I, 20230718A, which are very close to us for their
 ${\rm DM_{obs}}<1.5\;{\rm DM_{MW,\,ISM}}$.~After these robust cuts,
 the remaining 125 localized FRBs are used in this work.


 \begin{center}
 \begin{figure}[tb]
 \centering
 \vspace{-5mm}  
 \includegraphics[width=0.43\textwidth]{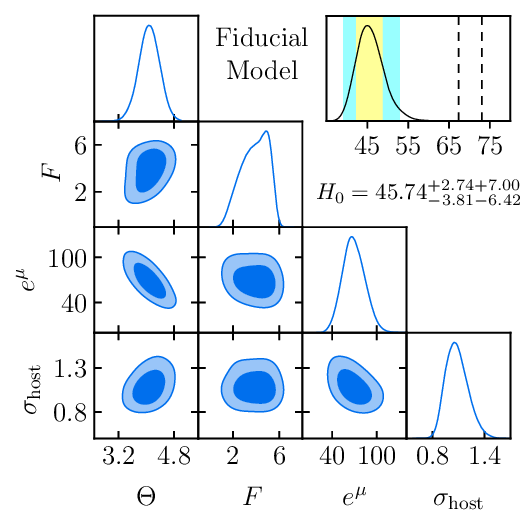}
 \vspace{-1mm}  
 \caption{\label{fig1} The $1\sigma$ and $2\sigma$ contours for all the
 free parameters of the fiducial model.~The top-right panel is
 the marginalized probability distribution of the Hubble constant $H_0$
 derived from $\Theta$ in Eq.~(\ref{eq4}) by using $f_{\rm IGM}=0.83$
 and $\Omega_b h^2=0.02237$.~The mean and $1\sigma$, $2\sigma$ intervals
 of $H_0$ are given numerically and also shown by the shaded
 regions.~$\Theta$ and $H_0$ are in units of $\rm km/s/Mpc$.~$e^\mu$ is
 in units of ${\rm pc\hspace{0.24em} cm^{-3}}$.~$H_0=67.36$ and $73.04\;
 {\rm km/s/Mpc}$ of Planck 2018 and SH0ES are also indicated by the vertical
 dashed lines.~See Sec.~\ref{sec2c} for details.}
 \end{figure}
 \end{center}


\vspace{-10mm} 


\subsection{Model comparison}\label{sec2b}

For model comparison, here we briefly introduce the Bayesian evidence
 and the information criteria. The Bayesian evidence is defined
 by~\cite{Kilbinger:2009by,Heavens:2017hkr} (see
 also e.g.~\cite{Jia:2025prq})
 \be{eq12}
 {\cal Z}=\int {\cal L}(\boldsymbol{\psi})\,\Phi(\boldsymbol{\psi})\,d
 \boldsymbol{\psi}\,,
 \ee
 where $\cal L$ is the likelihood function, $\Phi$ is the prior
 distribution, and $\boldsymbol{\psi}$ denotes the model parameters.~For
 model comparison, it is convenient to use the Bayes factor
 \be{eq13}
 {\cal B}_{12}={\cal Z}_1/{\cal Z}_2\,,\quad {\rm or~equivalently,}
 \quad\ln {\cal B}_{12}=\ln {\cal Z}_1 -\ln {\cal Z}_2\,,
 \ee
 where ${\cal Z}_1$ and ${\cal Z}_2$ are the Bayesian evidences for
 models $Q_1$ and $Q_2$, respectively.~If ${\cal B}_{12}$ is larger
 (smaller) than $1$, equivalently, if $\ln {\cal B}_{12}$ is positive
 (negative), model $Q_1$ ($Q_2$) is preferred over the other model.~The
 strength of evidence is indicated by the empirical ranges of
 $\left|\,\ln {\cal B}\,\right|$ summarized
 in Table~\ref{tab2}~\cite{Kilbinger:2009by,Heavens:2017hkr} (see also
 e.g.~\cite{Jia:2025prq}).~Note that one can compute the
 Bayesian evidence by using nested sampling (such as PolyChord,
 dynesty, MultiNest, nessai), or alternatively MCEvidence with
 MCMC chains~\cite{Heavens:2017afc,MCEvidence,MCEvimod}.~In this work,
 we use MCEvidence for convenience.

Additionally, some approximations of the Bayesian evidence such
 as the Akaike information criterion (AIC) and the Bayesian information
 criterion (BIC) have been extensively used
 in the literature.~The AIC is defined by~\cite{Akaike:1974}
 \be{eq14}
 {\rm AIC}=-2\ln {\cal L}_{\rm max}+2\kappa\,,
 \ee
 where ${\cal L}_{\rm max}$ is the maximum likelihood, and $\kappa$
 is the number of free model parameters.~For a Gaussian distribution,
 $\chi^2_{\rm min}=-2\ln {\cal L}_{\rm max}\,$.~The BIC is
 defined by~\cite{Schwarz:1978}
 \be{eq15}
 {\rm BIC}=-2\ln {\cal L}_{\rm max}+\kappa\ln N\,,
 \ee
 where $N$ is the number of data points.~Note that a smaller AIC or BIC
 indicates a better fit for the given model.~We can compare two models $Q_1$
 and $Q_2$ by calculating the differences in AIC and BIC, namely $\Delta
 {\rm AIC_{12}=AIC_1-AIC_2}$ and $\Delta {\rm BIC_{12}=BIC_1-
 BIC_2}$.~A negative (positive) $\Delta {\rm AIC_{12}}$ or
 $\Delta {\rm BIC_{12}}$ means a preference for model $Q_1$
 ($Q_2$).~The strength of evidence is indicated by the empirical ranges
 of $\left|\hspace{0.2mm}\Delta {\rm AIC}\hspace{0.3mm}\right|$
 or $\left|\hspace{0.2mm}\Delta {\rm BIC}\hspace{0.3mm}\right|$
 summarized in Table~\ref{tab2}~\cite{Perivolaropoulos:2022khd}
 (see also e.g.~\cite{Jia:2025prq}).~Comparing Eqs.~(\ref{eq14}) and
 (\ref{eq15}), it is more difficult to return a preference for a model
 when using $\Delta {\rm BIC}$ than $\Delta {\rm AIC}$ if $\ln N>2$,
 since the number of additional parameters in $\kappa$ is multiplied by
 $\ln N>2$ leading to a more rigorous penalty. In the following sections, we
 compare models using the Bayesian evidence and the information criteria
 AIC, BIC.


 \begin{center}
 \begin{figure}[tb]
 \centering
 \vspace{-5mm}  
 \includegraphics[width=0.43\textwidth]{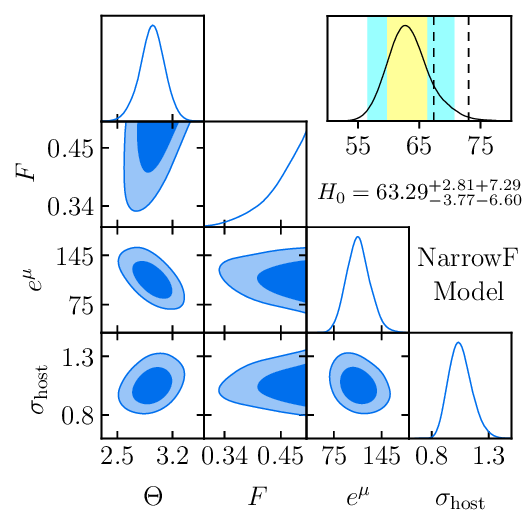}
 \vspace{-1mm}  
 \caption{\label{fig2} The same as in Fig.~\ref{fig1}, but for
 the narrow priors used by Macquart
 {\it et al.}~\cite{Macquart:2020lln}.~See Sec.~\ref{sec2c} for
 details.}
 \end{figure}
 \end{center}



 \begin{center}
 \begin{figure}[tb]
 \centering
 \vspace{-6.8mm}  
 \includegraphics[width=0.63\textwidth]{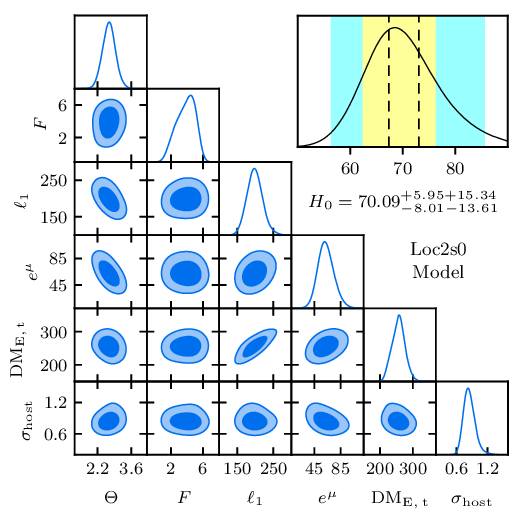}
 \vspace{-2mm}  
 \caption{\label{fig3} The same as in Fig.~\ref{fig1}, but
 for the Loc2s0 model.~See Sec.~\ref{sec3} for details.}
 \end{figure}
 \end{center}



 \begin{table}[tb]
 \renewcommand{\arraystretch}{1.2}
 \begin{center}
 \vspace{3.6mm}   
 \hspace{-1.5mm}  
 \begin{tabular}{lccccc}\hline\hline \\[-3.6mm]
   & Loc2s0 & Loc3s0 & Loc2s & Loc3s & Loclin \\[1mm] \hline \\[-3.6mm]
 $\Theta$ & $[\,0.01,\,5.0\,]$ & $[\,0.01,\,5.0\,]$ & $[\,0.01,\,5.0\,]$ & $[\,0.01,\,5.0\,]$ & $[\,0.01,\,5.0\,]$ \\
 $F$ & $[\,0.01,\,10.0\,]$ & $[\,0.01,\,10.0\,]$ & $[\,0.01,\,10.0\,]$ & $[\,0.01,\,10.0\,]$ & $[\,0.01,\,10.0\,]$ \\
 $\ell_0$ &  &  & $[\,-500,\,500\,]$ & $[\,-300,\,300\,]$ & $[\,-400,\,400\,]$ \\
 $\ell_1$ & $[\,0,\,500\,]$ & $[\,0,\,500\,]$ & $[\,0,\,500\,]$ & ~~~~$[\,-300,\,1000\,]$~~~~ & $[\,0,\,500\,]$ \\
 $\ell_2$ &  & $[\,0,\,1000\,]$ &  & $[\,0,\,1000\,]$ & \\
 $\rm DM_{E,\,t1}$~~~~ & $[\,0.01,\,1000\,]$ & ~~~~$[\,0.01,\,500\,]$~~~~ & $[\,0.01,\,1000\,]$ & $[\,0.01,\,500\,]$ & \\
 $S$ &   & $[\,0,\,500\,]$ &  & $[\,0.01,\,500\,]$ & \\
 $e^\mu$ & $[\,0.01,\,400\,]$ & $[\,0.01,\,400\,]$ & $[\,0.01,\,400\,]$ & $[\,0.01,\,400\,]$ & $[\,0.01,\,400\,]$ \\
 $\sigma_{\rm host}$ & $[\,0.01,\,4.0\,]$ & $[\,0.01,\,4.0\,]$ & $[\,0.01,\,4.0\,]$ & $[\,0.01,\,4.0\,]$ & $[\,0.01,\,4.0\,]$ \\[1mm]
 \hline\hline
 \end{tabular}
 \end{center}
 \vspace{-1.2mm}  
 \caption{\label{tabprior1} The uniform priors for all the free
 parameters of the $\ell$ models.~Note that $\rm DM_{E,\,t1}$ should
 be regarded as $\rm DM_{E,\,t}$ in the Loc2s0 and Loc2s models.~See
 Sec.~\ref{sec3} for details.}
 \end{table}


\vspace{-17mm} 


\section{Analysis}\label{sec3new}



\subsection{The fiducial model}\label{sec2c}

Now, we can constrain the model parameters by using the methodology
 and 125 localized FRBs given in Sec.~\ref{sec2a}.~First, we
 consider the fiducial model with the following uniform priors
 \be{eq16}
 \Theta\in [\,0.1,\,5.0\,],\quad F\in [\,0.01,\,10.0\,],\quad e^\mu
 \in [\,0.01,\,400\,],\quad \sigma_{\rm host}\in [\,0.01,\,4.0\,]\,.
 \ee
 In the rest of this work, the other models will be compared with this
 fiducial model.~Fitting to the data, we find the
 constraints on the free model parameters given by their means
 with $1\sigma$ uncertainties,
 \be{eq17}
 \Theta=4.08^{+0.30}_{-0.30}\,,\quad
 F=3.89^{+1.54}_{-0.82}\,,\quad
 e^\mu=69.19^{+14.10}_{-15.90}\,,
 \quad \sigma_{\rm host}=1.08^{+0.10}_{-0.15}\,.
 \ee
 We also present their contours in Fig.~\ref{fig1}.~As
 mentioned above, we adopt $\Omega_b h^2=0.02237$ from the Planck 2018
 result~\cite{Planck:2018vyg} and $f_{\rm IGM}=0.83$~\cite{Deng:2013aga,
 Yang:2016zbm,Gao:2014iva,Zhou:2014yta,Qiang:2019zrs,Qiang:2020vta,
 Qiang:2021bwb,Qiang:2021ljr,Guo:2022wpf,Guo:2023hgb} to derive the
 Hubble constant $H_0$ from $\Theta$ in Eq.~(\ref{eq4}), namely
 \be{eq18}
 H_0=45.74^{+2.74}_{-3.81}\,(1\sigma)\;{}^{+7.00}_{-6.42}\,(2\sigma)\;
 {}^{+10.91}_{-7.32}\,(3\sigma)~{\rm km/s/Mpc}\,.
 \ee
 Note that we do this easily by defining derived parameters in the MCMC
 codes e.g.~Cobaya or CosmoMC with GetDist.~The derived constraint on
 $H_0$ (the top-right panel of Fig.~\ref{fig1}) is in great tension with
 the ones from the CMB and SNIa (vertical dashed lines) at the $7\sim
 8\sigma$ level.

Note that we have adopted a loose prior $[\,0.01,\,10.0\,]$ for
 the parameter $F$ in the fiducial model, and all the model parameters
 (especially $F$) can be well constrained by the data in this case, as
 shown in Fig.~\ref{fig1}. In the literature, the effect of $F$ on $H_0$
 was discussed in e.g.~\cite{Baptista:2023uqu} and Sec.~4.4
 of~\cite{Xu:2025ddk}, and it was found that a small $F$ is required to
 obtain a Hubble constant $H_0$ consistent with the ones of Planck 2018
 and SH0ES. In fact, this is one of the key assumptions of Macquart
 {\it et al.}~\cite{Macquart:2020lln} which adopted a narrow prior for
 $F$ with a fairly small upper bound $0.5$.~Here we also consider the
 same narrow priors as in Extended Data Table~2 of Macquart
 {\it et al.}~\cite{Macquart:2020lln}, namely


 \begin{center}
 \begin{figure}[tb]
 \centering
 \vspace{-8mm}  
 \includegraphics[width=0.86\textwidth]{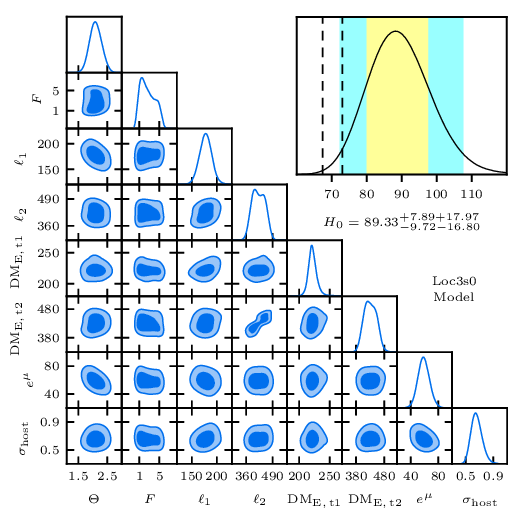}
 \vspace{-1mm}  
 \caption{\label{fig4} The same as in Fig.~\ref{fig1}, but
 for the Loc3s0 model.~See Sec.~\ref{sec3} for details.}
 \end{figure}
 \end{center}


\vspace{-14.5mm} 

 \be{eq19}
 \Theta\in [\,0.1,\,5.0\,],\quad F\in [\,0.011,\,0.5\,],\quad
 e^\mu \in [\,20,\,200\,],\quad \sigma_{\rm host}\in [\,0.2,\,2.0\,]\,.
 \ee
 They are all much narrower than the ones of our fiducial model
 in Eq.~(\ref{eq16}).~We refer to this as the NarrowF model.~Fitting to
 the data, we find the constraints on the free model parameters, namely
 \be{eq20}
 \Theta=2.94^{+0.16}_{-0.15}\,,\quad
 F=0.45^{+0.05}_{-0.01}\,,\quad
 e^\mu=110.81^{+14.87}_{-19.09}\,,
 \quad \sigma_{\rm host}=1.05^{+0.08}_{-0.12}\,,
 \ee
 and the derived Hubble constant
 \be{eq21}
 H_0=63.29^{+2.81}_{-3.77}\,(1\sigma)\;{}^{+7.29}_{-6.60}\,(2\sigma)\;
 {}^{+10.97}_{-7.81}\,(3\sigma)~{\rm km/s/Mpc}\,.
 \ee
 We also present the results in Fig.~\ref{fig2}.~It is worth noting that
 $H_0$ was not explicitly constrained in Macquart {\it et
 al.}~\cite{Macquart:2020lln}, where only the constraint on the
 combination $\Omega_b h_{70}$ (in which $h_{70}=H_0/(70\;{\rm
 km/s/Mpc})$) was given.~Additionally, only 8 localized FRBs was used by
 Macquart {\it et al.}~\cite{Macquart:2020lln}, while 125 localized
 FRBs are used in the present work.~This also makes a difference in the
 derived $H_0$.~Clearly, in the NarrowF model, $H_0$ given by
 Eq.~(\ref{eq21}) can be consistent with the ones of Planck 2018 and
 SH0ES at the $2\sigma$ and $3\sigma$ levels, respectively.~However, we
 note that the marginalized probability distribution of $F$ is skew and
 $F$ cannot be constrained from the right hand side, as shown
 in Fig.~\ref{fig2} (see also e.g.~Fig.~5 of~\cite{Zhuge:2025urk} and
 Extended Data Fig.~5 of~\cite{Macquart:2020lln}).~The fairly low upper
 bound $0.5$ of the narrow prior for $F$ prevents it from taking larger
 values.~If $F$ can be larger (e.g.~$F>1\sim 4$), $H_0$ becomes
 significantly inconsistent with the ones of Planck 2018 and SH0ES, as shown
 by the fiducial model.~In fact, $F\simeq 0.45$ of the NarrowF model is
 far outside $3\sigma$ region of the fiducial model (n.b.~Eq.~(\ref{eq17})).

To determine which model is preferred, we compute $\ln \cal B$,
 $\Delta \rm AIC$ and $\Delta \rm BIC$ of the NarrowF model relative to
 the fiducial model, and find
 \be{eq22}
 \ln {\cal B}=-6.68\,,
 \quad\Delta {\rm AIC}=13.43\,,\quad\Delta {\rm BIC}=13.43\,.
 \ee
 Thus, the fiducial model is strongly preferred over the NarrowF model
 in terms of the Bayesian evidence and the information criteria AIC, BIC
 (n.b.~Table~\ref{tab2}).~So, the great Hubble tension between FRBs,
 Planck 2018 and SH0ES found in the fiducial model should be
 taken seriously.


 \begin{table}[tb]
 \renewcommand{\arraystretch}{1.2}
 \begin{center}
 \vspace{-4mm}   
 \hspace{-1.5mm}  
 \begin{tabular}{lccccc}\hline\hline \\[-3.6mm]
   & Loc2s0 & Loc3s0 & Loc2s & Loc3s & Loclin \\[1mm] \hline \\[-3.6mm]
 $\Theta$ & $2.68^{+0.26}_{-0.27}$ & $2.10^{+0.20}_{-0.22}$ & $2.70^{+0.28}_{-0.30}$ & $2.04^{+0.21}_{-0.22}$ & $2.39^{+0.22}_{-0.22}$ \\
 $F$ & $3.86^{+1.41}_{-1.00}$ & $2.71^{+1.12}_{-1.98}$ & $2.91^{+1.76}_{-2.53}$ & $2.68^{+1.12}_{-2.14}$ & $4.49^{+0.96}_{-0.58}$ \\
 $\ell_0$ &   &   & $-61.90^{+75.32}_{-35.50}$ & ~~~~$-62.28^{+67.44}_{-36.91}$~~~~ & $-391.20^{+1.43}_{-8.80}$ \\
 $\ell_1$ & $198.49^{+20.41}_{-21.36}$ & ~~~~$177.50^{+11.69}_{-11.72}$~~~~ & $195.63^{+65.10}_{-61.12}$ & $113.51^{+70.56}_{-39.95}$ & $99.96^{+3.12}_{-1.85}$ \\
 $\ell_2$ &  & $421.23^{+32.95}_{-34.18}$ &   & $355.09^{+77.18}_{-53.36}$ & \\
 $\rm DM_{E,\,t1}$~~~~ & $256.14^{+19.76}_{-19.81}$ & $222.22^{+3.65}_{-7.04}$ & $334.77^{+78.73}_{-80.82}$ & $221.28^{+5.59}_{-7.21}$ & \\
 $S$ &   & $208.25^{+21.72}_{-23.37}$ &   & $205.86^{+21.13}_{-22.79}$ & \\
 $e^\mu$ & $62.65^{+10.95}_{-13.20}$ & $58.80^{+7.82}_{-8.53}$ & $149.04^{+48.33}_{-100.92}$ & $129.69^{+39.89}_{-73.82}$ & $28.56^{+8.13}_{-12.38}$ \\
 $\sigma_{\rm host}$ & $0.85^{+0.09}_{-0.12}$ & $0.66^{+0.07}_{-0.09}$ & $0.53^{+0.17}_{-0.26}$ & $0.39^{+0.12}_{-0.20}$ & $1.23^{+0.24}_{-0.32}$ \\[1mm] \hline \\[-3.6mm]
 $\rm DM_{E,\,t2}$ &   & $430.47^{+22.60}_{-23.30}$ &  & $427.15^{+20.92}_{-23.11}$ & \\
 $H_0$~($1\sigma$) & $70.09^{+5.95}_{-8.01}$ & $89.33^{+7.89}_{-9.72}$ & $69.57^{+6.49}_{-8.31}$ & $92.00^{+8.51}_{-10.86}$ & $78.36^{+5.87}_{-8.09}$ \\
 $H_0$~($2\sigma$) & $70.09^{+15.34}_{-13.61}$ & $89.33^{+17.97}_{-16.80}$ & $69.57^{+15.65}_{-14.42}$ & $92.00^{+20.42}_{-18.89}$ & $78.36^{+14.75}_{-13.27}$ \\
 $H_0$~($3\sigma$) & $70.09^{+22.33}_{-15.82}$ & $89.33^{+25.51}_{-21.00}$ & $69.57^{+24.66}_{-16.18}$ & $92.00^{+32.35}_{-21.09}$ & $78.36^{+21.89}_{-15.87}$ \\[1mm]
 \hline\hline
 \end{tabular}
 \end{center}
 \vspace{-1.5mm}  
 \caption{\label{tabcons1} The means and $1\sigma$ uncertainties for all
 the free parameters and the derived parameter $\rm DM_{E,\,t2}$, as well as
 the means and $1-3\sigma$ uncertainties for the derived parameter $H_0$
 (last three rows) of the $\ell$ models.~Note that $\rm DM_{E,\,t1}$
 should be regarded as $\rm DM_{E,\,t}$ in the Loc2s0 and Loc2s
 models.~See Sec.~\ref{sec3} for details.}
 \end{table}



 \begin{center}
 \begin{figure}[tb]
 \centering
 \vspace{-6.4mm}  
 \includegraphics[width=0.75\textwidth]{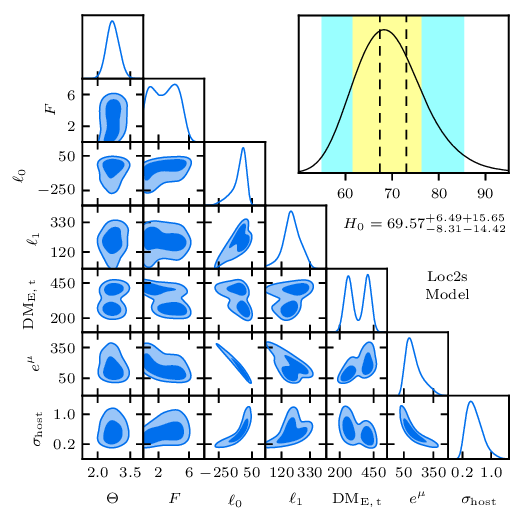}
 \vspace{-2mm}  
 \caption{\label{fig5} The same as in Fig.~\ref{fig1}, but
 for the Loc2s model.~See Sec.~\ref{sec3} for details.}
 \end{figure}
 \end{center}



 \begin{center}
 \begin{figure}[tb]
 \centering
 \vspace{-7.5mm}  
 \includegraphics[width=0.95\textwidth]{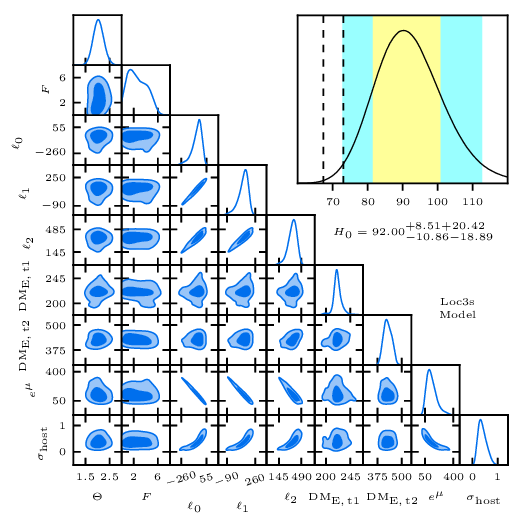}
 \vspace{-2mm}  
 \caption{\label{fig6} The same as in Fig.~\ref{fig1}, but
 for the Loc3s model.~See Sec.~\ref{sec3} for details.}
 \end{figure}
 \end{center}



 \begin{center}
 \begin{figure}[tb]
 \centering
 \vspace{-7.5mm} 
 \includegraphics[width=0.63\textwidth]{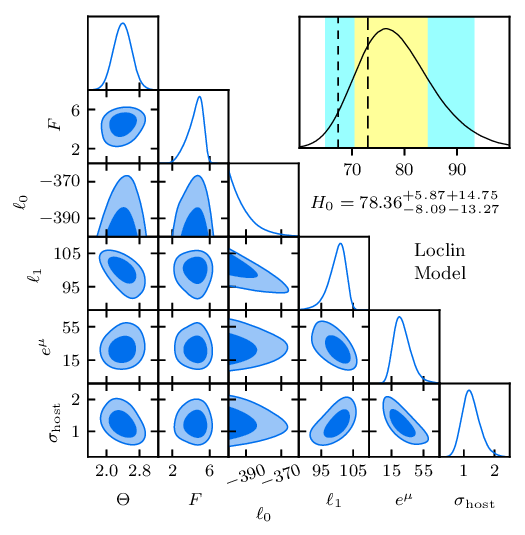}
 \vspace{-2.45mm}  
 \caption{\label{fig7} The same as in Fig.~\ref{fig1}, but
 for the Loclin model.~See Sec.~\ref{sec3} for details.}
 \end{figure}
 \end{center}


\vspace{-25.4mm} 



\subsection{Generalized distributions of $\bf DM_{host}$
 with varying location $\boldsymbol{\ell}$}\label{sec3}

Instead of modifying $\sigma_\Delta=Fz^{-0.5}$ in the distribution of
 $\rm DM_{IGM}$ as in e.g.~\cite{Zhuge:2025urk}, we study the effect of
 the generalized distribution of $\rm DM_{host}$ on $H_0$ in the present
 work.~Note that the log-normal distribution Eq.~(\ref{eq8}) used in the
 fiducial methodology is not the general form.~According to
 e.g.~\cite{lognorm}, one can shift and/or scale the standard log-normal
 distribution by using the location and scale parameters $\ell$ and
 $e^\mu$ (both in units of ${\rm pc\hspace{0.24em} cm^{-3}}$),
 and hence the most general log-normal distribution of $\rm
 DM_{host}$ is given by
 \bea
 &&\hspace{-15mm}P_{\rm host}({\rm DM_{host}}\,|\,\ell,\,
 \mu,\,\sigma_{\rm host})=\nonumber\\[0.1mm]
 &&\quad\frac{1}{\sigma_{\rm host}\left({\rm DM_{host}}-\ell\right)
 \sqrt{2\pi}}\cdot\exp\left[-\frac{\left(\,\log\left({\rm DM_{host}}-
 \ell\right)-\mu\right)^2}{2\sigma_{\rm host}^2}\,\right]\,
 \quad\quad {\rm for}\quad {\rm DM_{host}}>\ell\,,\label{eq23}
 \eea
 and $P_{\rm host}({\rm DM_{host}})=0$ for ${\rm DM_{host}}\leq\ell$.~In
 the fiducial methodology proposed by Macquart {\it et
 al.}~\cite{Macquart:2020lln}, a zero location parameter $\ell=0$ has
 been implicitly used.~In fact, a non-zero location parameter $\ell$
 shifts the log-normal distribution of $\rm DM_{host}$ as a whole.~If
 $\ell>0$, $\rm DM_{host}$ will be shifted toward larger values, or
 equivalently $\rm DM_{IGM}$ will be shifted toward smaller values for
 fixed ${\rm DM_E=DM_{IGM}+DM_{host}}/(1+z)$, and hence $\Theta$ will be
 smaller (n.b.~Eq.~(\ref{eq4})).~A smaller $\Theta\propto
 (\Omega_b h^2)\,f_{\rm IGM}/H_0$ leads to a larger Hubble constant
 $H_0$ for fixed $\Omega_b h^2$ and $f_{\rm IGM}$.~This physical picture
 is fairly clear, in which $H_0$ from localized FRBs can be
 consistent with the ones of Planck 2018 and SH0ES naturally.


 \begin{table}[tb]
 \renewcommand{\arraystretch}{1.7}
 \begin{center}
 \vspace{3.6mm}   
 \hspace{-1.5mm}  
 \begin{tabular}{ccccc|ccccc}\hline\hline
 Model & $\ln \cal B$ & $\Delta \rm AIC$ & $\Delta \rm BIC$ & & \hspace{1.2mm} & Model & $\ln \cal B$ & $\Delta \rm AIC$ & $\Delta \rm BIC$ \\ \hline
 Loc2s0 & 37.23 & $-86.28$ & $-80.62$ & & & Mu2s & 29.32 & $-70.94$ & $-65.28$ \\[-2mm]
 ~Loc3s0~ & ~74.28~ & ~$-170.23$~ & ~$-158.92$~ & & & ~Mu3s~ & ~~50.93~~ & ~$-127.09$~ & ~$-115.78$~ \\[-2mm]
 Loc2s & 34.49 & $-84.04$ & $-75.56$ & & & Mulin & 39.14 & $-90.09$ & $-87.26$ \\[-2mm]
 Loc3s & 71.74 & $-168.88$ & $-154.74$ & & & Fiducial & 0 & 0 & 0 \\[-2mm]
 Loclin & 49.12 & $-112.59$ & $-104.11$ & & & NarrowF & $-6.68$ & 13.43 & 13.43 \\
 \hline\hline
 \end{tabular}
 \end{center}
 \vspace{-1mm}  
 \caption{\label{tab3} The Bayesian evidences and
 the information criteria AIC, BIC for all models relative
 to the fiducial model (n.b.~Table~\ref{tab2}).}
 \end{table}


It is worth noting that one cannot make $\ell$ disappear by redefining
 ${\rm DM_{host}^\prime=DM_{host}}-\ell$ in $P_{\rm host}$.~With this
 new $\rm DM_{host}^\prime$, one should also change the
 integral variable $\rm DM_{host}$ in Eq.~(\ref{eq9}) to $\rm
 DM_{host}^\prime$ accordingly, and then the term ${\rm DM_E-
 DM_{host}}/(1+z)$ in $P_{\rm IGM}$ becomes ${\rm DM_E}-
 ({\rm DM_{host}^\prime}+\ell)/(1+z)$.~So, $\ell$ does not disappear but
 changes its position in the likelihood $\cal L$,
 and an $\ell>0$ still leads to a larger $H_0$.

The naive idea to adopt a universal $\ell=const.$ does not
 work well.~The key point is that ${\rm DM_{host}}-\ell>0$ is required
 in Eq.~(\ref{eq23}).~Thus, $\ell<{\rm DM_{host}}=(1+z)({\rm DM_E-DM_{IGM}})
 \leq\left(1+z\right){\rm DM_E}$.~If $\ell=const.$ is universal for all
 localized FRBs, it should be less than the smallest $\left(1+z_i\right){\rm
 DM}_{{\rm E},\,i}$ of all localized FRBs. This will force $\ell=const.$ to
 be fairly small (close to $0$), not so different from the fiducial model.

Note that if $\ell\leq 0$, the lower limit of integral in
 Eq.~(\ref{eq9}) is still $0$ since ${\rm DM_{host}}>0$.~But
 theoretically the lower limit of integral in Eq.~(\ref{eq9}) should be
 changed to $\ell$ if $\ell>0$, due to the requirement ${\rm DM_{host}}>
 \ell$.~In practice, however, we can still use a zero lower limit of
 integral in Eq.~(\ref{eq9}), since the integral from $0$ to $\ell$ is
 actually zero due to the fact that the log-normal PDF $\!f(x,\,s,\,
 loc,\,scale)$ should be $0$ for $x\leq loc$ (n.b.~the line
 below Eq.~(\ref{eq23})).~Even so, on the other hand, $\ell<\left(1+z\right)
 {\rm DM_E}$ is still required, otherwise the whole integral in
 Eq.~(\ref{eq9}) will be zero since $P_{\rm IGM}({\rm DM_E-DM_{host}}/(1+z))
 =0$ for ${\rm DM_{host}}>\ell\geq\left(1+z\right){\rm DM_E}$
 (n.b.~$P_{\rm IGM}(\Delta)=0$ for $\Delta\leq 0$ below
 Eq.~(\ref{eq6})), and then the total likelihood ${\cal L}=0$
 or $\log{\cal L}$ diverges.~In summary, $\ell<{\rm DM_{host}}
 \leq\left(1+z\right){\rm DM_E}$ is required, but technically
 the lower limit of integral in Eq.~(\ref{eq9}) can still
 be $0$ in practice for convenience.

Since a universal $\ell=const.$ does not work well due to the FRBs with
 low values of $\left(1+z\right){\rm DM_E}$ as mentioned above, it is
 reasonable to accommodate these FRBs with a step-like $\ell$, namely
 \be{eq24}
 \ell=
 \begin{cases}
 \;\ell_0=0 \quad & {\rm if}\hspace{2mm} {\rm DM}_{{\rm E},\,i}
 <{\rm DM_{E,\,t}}\,,\\[1.6mm]
 \;\ell_1 & {\rm if}\hspace{2mm} {\rm DM}_{{\rm E},\,i}\geq
 {\rm DM_{E,\,t}}\,.
 \end{cases}
 \ee
 We label this model as Loc2s0, with two new model parameters $\ell_1$ and
 ${\rm DM_{E,\,t}}$ (both in units of ${\rm pc\hspace{0.24em} cm^{-3}}$). It
 converges to the fiducial model for the FRBs with low ${\rm DM}_{{\rm E},\,
 i}$.~The uniform priors for its free model parameters are presented in
 Table~\ref{tabprior1}, and the resultant constraints on the free model
 parameters and the derived parameter $H_0$ are shown in Fig.~\ref{fig3}
 and Table~\ref{tabcons1}.~The derived constraints on $H_0$ are
 consistent with the ones of Planck 2018 and SH0ES well within
 the $1\sigma$ region.~When comparing the Loc2s0 and fiducial models
 (n.b.~Table~\ref{tab3}), we find an overwhelming preference for the
 Loc2s0 model from the Bayes factor and AIC, BIC.~So, the generalized
 distribution of $\rm DM_{host}$ with $\ell$ works very well.

It is natural to go further by considering a three-step $\ell$, namely
 \be{eq29}
 \ell=
 \begin{cases}
 \;\ell_0=0 \quad & {\rm if}\hspace{2mm} {\rm DM}_{{\rm E},\,i}
 <{\rm DM_{E,\,t1}}\,,\\[1.2mm]
 \;\ell_1 & {\rm if}\hspace{2mm} {\rm DM_{E,\,t1}}\leq {\rm
 DM}_{{\rm E},\,i}<{\rm DM_{E,\,t2}}\,,\\[1.2mm]
 \;\ell_2 & {\rm if}\hspace{2mm} {\rm DM}_{{\rm E},\,i}\geq
 {\rm DM_{E,\,t2}}\,.
 \end{cases}
 \ee
 We label this model as Loc3s0, with four new model parameters $\ell_1$,
 $\ell_2$, ${\rm DM_{E,\,t1}}$ and ${\rm DM_{E,\,t2}}$ (all in units
 of ${\rm pc\hspace{0.24em} cm^{-3}}$).~Note that for convenience in setting
 the priors properly, we alternatively consider a free model parameter
 $S\geq 0$, and then regard ${\rm DM_{E,\,t2}}={\rm DM_{E,\,t1}}+S$ as
 a derived parameter.~This is just a technical trick to ensure
 ${\rm DM_{E,\,t2}}\geq {\rm DM_{E,\,t1}}$.

In the above cases, $\ell_0=0$ has been set.~We can let it be free, and
 consider
 \be{eq34}
 \ell=
 \begin{cases}
 \;\ell_0 \quad & {\rm if}\hspace{2mm} {\rm DM}_{{\rm E},\,i}
 <{\rm DM_{E,\,t}}\,,\\[1.5mm]
 \;\ell_1 & {\rm if}\hspace{2mm} {\rm DM}_{{\rm E},\,i}\geq
 {\rm DM_{E,\,t}}\,.
 \end{cases}
 \ee
 We label this model as Loc2s.~Similarly, we also consider the Loc3s
 model, in which
 \be{eq39}
 \ell=
 \begin{cases}
 \;\ell_0 \quad & {\rm if}\hspace{2mm} {\rm DM}_{{\rm E},\,i}
 <{\rm DM_{E,\,t1}}\,,\\[1.2mm]
 \;\ell_1 & {\rm if}\hspace{2mm} {\rm DM_{E,\,t1}}\leq {\rm
 DM}_{{\rm E},\,i}<{\rm DM_{E,\,t2}}\,,\\[1.2mm]
 \;\ell_2 & {\rm if}\hspace{2mm} {\rm DM}_{{\rm E},\,i}\geq
 {\rm DM_{E,\,t2}}\,,
 \end{cases}
 \ee
 and ${\rm DM_{E,\,t2}}={\rm DM_{E,\,t1}}+S$ is a derived parameter,
 while $S\geq 0$ is a free model parameter.


 \begin{center}
 \begin{figure}[tb]
 \centering
 \vspace{-6mm}  
 \includegraphics[width=0.49\textwidth]{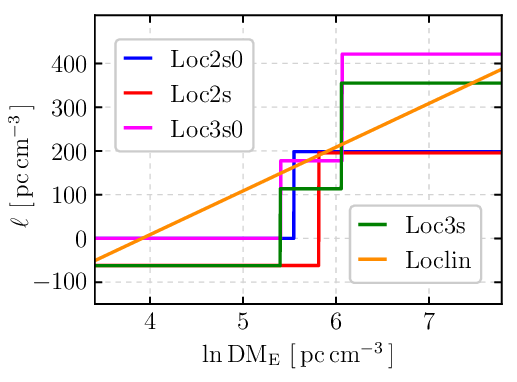}
 \vspace{-1.5mm}  
 \caption{\label{fig8} $\ell$ versus $\ln {\rm DM_E}$ for all the $\ell$
 models.~See Sec.~\ref{sec3} for details.}
 \end{figure}
 \end{center}



 \begin{table}[tb]
 \renewcommand{\arraystretch}{1.2}
 \begin{center}
 \vspace{3.6mm}   
 \hspace{-1.5mm}  
 \begin{tabular}{lccc}\hline\hline \\[-3.6mm]
   & Mu2s & Mu3s & Mulin \\[1mm] \hline \\[-3.6mm]
 $\Theta$ & $[\,0.01,\,5.0\,]$ & $[\,0.01,\,5.0\,]$ & $[\,0.01,\,5.0\,]$ \\
 $F$ & $[\,0.01,\,10.0\,]$ & $[\,0.01,\,10.0\,]$ & $[\,0.01,\,10.0\,]$ \\
 $\mu_0$ & $[\,0.01,\,500\,]$ & $[\,0.01,\,500\,]$ & $[\,-1000,\,200\,]$ \\
 $\mu_1$ & $[\,0.01,\,500\,]$ & $[\,0.01,\,500\,]$ & $[\,-10,\,500\,]$ \\
 $\mu_2$ &  & ~~~~~$[\,0.01,\,1000\,]$~~~~~ & \\
 $\rm DM_{E,\,t1}$~~~~ & $[\,0.01,\,1000\,]$ & $[\,0.01,\,500\,]$ & \\
 $S$ &   & $[\,0.01,\,500\,]$  & \\
 $\sigma_{\rm host}$ & $[\,0.01,\,4.0\,]$ & $[\,0.01,\,4.0\,]$ & $[\,0.01,\,4.0\,]$ \\[1mm]
 \hline\hline
 \end{tabular}
 \end{center}
 \vspace{-1.5mm}  
 \caption{\label{tabprior2} \,The uniform priors for all the
 free parameters of the $e^\mu$ models.~Note that $\rm DM_{E,\,t1}$
 should be regarded as $\rm DM_{E,\,t}$ in the Mu2s model.~See
 Sec.~\ref{sec4} for details.}
 \end{table}


\vspace{-9mm} 

The uniform priors for the free parameters of the Loc3s0, Loc2s, Loc3s
 models are given in Table~\ref{tabprior1}, and the resultant constraints on
 the free model parameters and the derived parameters $H_0$,
 $\rm DM_{E,\,t2}$ are shown in Figs.~$\ref{fig4}-\ref{fig6}$
 and Table~\ref{tabcons1}.~The derived constraints on $H_0$ for these models
 are consistent with the ones of Planck 2018 and SH0ES within
 the $1\sigma$ or $2\sim 3\sigma$ regions.~When comparing the Loc3s0, Loc2s,
 Loc3s and fiducial models (n.b.~Table~\ref{tab3}), we find an
 overwhelming preference for all these step-like $\ell$ models from
 the Bayes factor and AIC, BIC.~Note that the Loc2s0 model is mildly
 preferred over the Loc2s model despite the latter having an additional free
 parameter $\ell_0$.~Indeed, $\ell_0=0$ is still in the $1\sigma$ region
 of $\ell_0$ for the Loc2s model (n.b.~Table~\ref{tabcons1}).~The Loc3s0 and
 Loc3s models are also in such a similar situation. In general, the
 models with more steps in $\ell$ are preferred when comparing
 the Bayesian evidence, AIC and BIC (n.b.~Table~\ref{tab3}), and there is no
 preference for $\ell_0\not=0$.

Because the form of $\ell$ is not well known, we have thus far
 approximated it with step-like functions. Alternatively, we
 might consider its Taylor expansion with respect to $\rm DM_E$ up to
 first order, namely $\ell=\ell_0+\ell_1\, {\rm DM_E}$.~But $\rm DM_E$
 of FRBs are large numbers spanning three orders of magnitude ${\cal O}
 (10\sim 10^3)$. In this case, the higher orders cannot be dropped. Usually,
 one can smooth it by using logarithm. Instead, we consider
 $\ell$'s Taylor expansion with respect to $\ln \rm DM_E$ up to
 first order, namely
 \be{eq44}
 \ell=\ell_0+\ell_1\ln {\rm DM_E}\,.
 \ee
 We label this model as Loclin.~The uniform priors for its free model
 parameters are presented in Table~\ref{tabprior1}, and the resultant
 constraints on the free model parameters and the derived
 parameter $H_0$ are shown in Fig.~\ref{fig7} and Tables~\ref{tabcons1}.~The
 derived constraints on $H_0$ are consistent with the ones of Planck
 2018 and SH0ES within $1\sim 2\sigma$ regions.~When comparing
 the Loclin and fiducial models (n.b.~Table~\ref{tab3}), we find an
 overwhelming preference for the Loclin model from the Bayes factor and
 AIC, BIC.

It is of interest to compare all the $\ell$ models.~In Fig.~\ref{fig8},
 we present $\ell$ versus $\ln {\rm DM_E}$ for all the $\ell$
 models, while the model parameters take their mean values obtained from
 the data.~We find that $\ell_0\sim -60$ or $0$, and $\ell_1\sim 200$ or
 $100$ for all step-like $\ell$ models, while the transition(s) might
 happen at $\ln {\rm DM_E}\sim 5.5$.~The $\ell$ of Loclin model crosses
 almost all steps of $\ell_i$, as shown in Fig.~\ref{fig8}, and hence
 this linear $\ell$ model is at least a decent enough approximation of
 the unknown $\ell$ function.~Notice that the common feature of all the
 $\ell$ models is that $\ell$ converges to a small value (around $0$) to
 accommodate the localized FRBs at low $\rm DM_E$ as in the fiducial
 model, and then $\ell$ becomes larger at high $\rm DM_E$ which shifts
 $H_0$ to larger values.~\hspace{0.15em}The physical picture is quite
 clear.~\hspace{0.15em}In Table~\ref{tab3}, we also summarize
 the Bayesian evidences and the information criteria AIC, BIC for all
 the $\ell$ models relative to the fiducial model.~One can easily see
 that $\rm Loc3s0>Loc3s\gg Loclin\gg Loc2s0>Loc2s\gg Fiducial$.~The
 Hubble constants $H_0$ can be consistent with the ones of Planck 2018
 and SH0ES in all the $\ell$ models.


 \begin{center}
 \begin{figure}[tb]
 \centering
 \vspace{-6mm}  
 \includegraphics[width=0.64\textwidth]{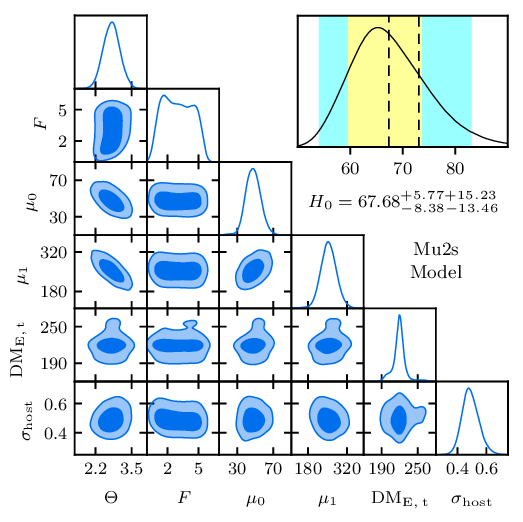}
 \vspace{-1mm}  
 \caption{\label{fig9} The same as in Fig.~\ref{fig1}, but
 for the Mu2s model.~See Sec.~\ref{sec4} for details.}
 \end{figure}
 \end{center}



 \begin{center}
 \begin{figure}[tb]
 \centering
 \vspace{-8mm}  
 \includegraphics[width=0.86\textwidth]{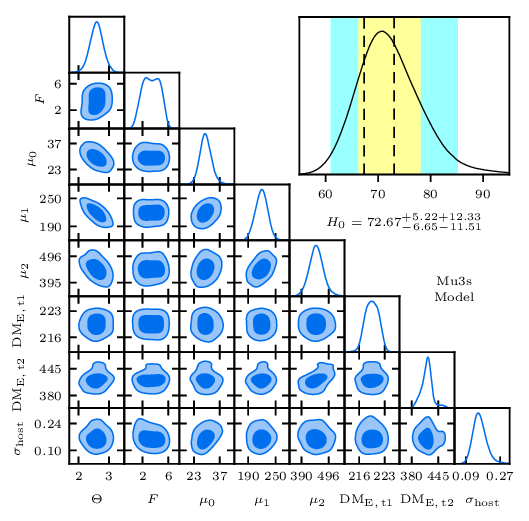}
 \vspace{-1mm}  
 \caption{\label{fig10} The same as in Fig.~\ref{fig1}, but
 for the Mu3s model.~See Sec.~\ref{sec4} for details.}
 \end{figure}
 \end{center}



 \begin{center}
 \begin{figure}[tb]
 \centering
 \vspace{-6mm}  
 \includegraphics[width=0.54\textwidth]{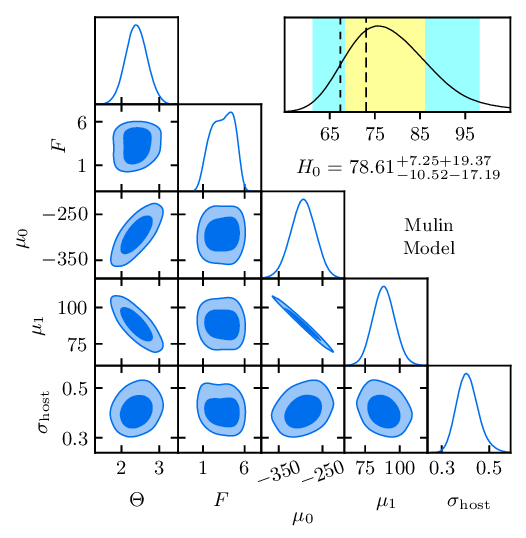}
 \vspace{-1mm}  
 \caption{\label{fig11} The same as in Fig.~\ref{fig1}, but
 for the Mulin model.~See Sec.~\ref{sec4} for details.}
 \end{figure}
 \end{center}



 \begin{table}[tb]
 \renewcommand{\arraystretch}{1.2}
 \begin{center}
 \vspace{4.5mm}   
 \hspace{-1.5mm}  
 \begin{tabular}{lccc}\hline\hline \\[-3.6mm]
   & Mu2s & Mu3s & Mulin \\[1mm] \hline \\[-3.6mm]
 $\Theta$ & $2.78^{+0.30}_{-0.30}$ & $2.57^{+0.21}_{-0.22}$ & $2.40^{+0.28}_{-0.28}$ \\
 $F$ & $3.02^{+1.44}_{-1.64}$ & $3.29^{+1.53}_{-1.47}$ & $3.37^{+1.54}_{-1.31}$ \\
 $\mu_0$ & $47.68^{+7.82}_{-8.40}$ & $29.70^{+2.56}_{-3.71}$ & $-294.19^{+28.46}_{-27.86}$ \\
 $\mu_1$ & $253.97^{+27.55}_{-27.70}$ & ~~~~~~~$219.53^{+12.56}_{-12.54}$~~~~~~~ & $88.51^{+8.01}_{-7.97}$ \\
 $\mu_2$ &  & $449.02^{+25.48}_{-26.57}$ & \\
 $\rm DM_{E,\,t1}$~~~~~~ & $217.96^{+10.33}_{-9.27}$ & $219.01^{+2.62}_{-1.88}$ & \\
 $S$ &   & $198.19^{+11.93}_{-12.09}$ & \\
 $\sigma_{\rm host}$ & $0.49^{+0.05}_{-0.06}$ & $0.17^{+0.03}_{-0.04}$ & $0.41^{+0.04}_{-0.05}$ \\[1mm] \hline \\[-3.6mm]
 $\rm DM_{E,\,t2}$ &   & $417.20^{+11.86}_{-11.22}$ & \\
 $H_0$~($1\sigma$) & $67.68^{+5.77}_{-8.38}$ & $72.67^{+5.22}_{-6.65}$ & $78.61^{+7.25}_{-10.52}$ \\
 $H_0$~($2\sigma$) & $67.68^{+15.23}_{-13.46}$ & $72.67^{+12.33}_{-11.51}$ & $78.61^{+19.37}_{-17.19}$ \\
 $H_0$~($3\sigma$) & $67.68^{+24.85}_{-15.26}$ & $72.67^{+18.47}_{-13.67}$ & $78.61^{+31.40}_{-19.19}$ \\[1mm]
 \hline\hline
 \end{tabular}
 \end{center}
 \vspace{-1mm}  
 \caption{\label{tabcons2} The means and $1\sigma$ uncertainties for all
 the free parameters and the derived parameter $\rm DM_{E,\,t2}$, as well as
 the means and $1-3\sigma$ uncertainties for the derived parameter $H_0$
 (last three rows) of the $e^\mu$ models.~Note that $\rm DM_{E,\,t1}$
 should be regarded as $\rm DM_{E,\,t}$ in the Mu2s model.~See
 Sec.~\ref{sec4} for details.}
 \end{table}


\vspace{-25.9mm} 



\subsection{Generalized distributions of $\bf DM_{host}$ with
 varying scale $\boldsymbol{e^\mu}$}\label{sec4}

As discussed in the beginning of Sec.~\ref{sec3}, the key to make the Hubble
 constant $H_0$ consistent with the ones of Planck 2018 and SH0ES is
 to shift $\rm DM_{host}$ toward larger values.~In Sec.~\ref{sec3}, we
 use the location parameter $\ell$ to this end, which shifts
 the log-normal distribution of $\rm DM_{host}$ as a whole.~In this
 section, we try an alternative way.~\hspace{0.1em}Let us come back to
 Eq.~(\ref{eq8}) without the location parameter $\ell$ (namely
 $\ell=0$), and we note that the $\rm DM_{host}$ distribution
 in Eq.~(\ref{eq8}) has a median value of $e^\mu$ (in units of
 ${\rm pc\hspace{0.24em} cm^{-3}}$)~\cite{Macquart:2020lln,lognorm},
 which plays the role of scale parameter.~If the median value $e^\mu$
 of the log-normal $\rm DM_{host}$ distribution is shifted toward higher
 values, this also produces larger values of $\rm DM_{host}$ from this
 scaled log-normal distribution.~\hspace{0.1em}The difference is that
 $\ell$ directly shifts the log-normal distribution as a whole, while
 $e^\mu$ only shifts the median value of the log-normal distribution.~Hence,
 it works indirectly, and its effect could be slightly weaker than the
 effects of modifying $\ell$.~Nevertheless, the scale parameter $e^\mu$
 can also make $H_0$ consistent with the ones of Planck 2018
 and SH0ES.~The physical picture is clear.

Similar to the case of $\ell$, one cannot make $e^\mu$ disappear by
 redefining ${\rm DM_{host}^\prime=DM_{host}}/e^\mu$ in $P_{\rm host}$.~With
 this new $\rm DM_{host}^\prime$, one should also change the integral
 variable $\rm DM_{host}$ in Eq.~(\ref{eq9}) to $\rm DM_{host}^\prime$
 accordingly, and then the term ${\rm DM_E-DM_{host}}/(1+z)$ in
 $P_{\rm IGM}$ becomes ${\rm DM_E}-e^\mu\,{\rm DM_{host}^\prime}/(1+z)$.~So,
 $e^\mu$ does not disappear but changes its position in the likelihood
 $\cal L$, and a larger $e^\mu$ still leads to a higher $H_0$.

In the fiducial model, $e^\mu$ is a universal constant for all
 localized FRBs.~We want to shift $e^\mu$ toward larger values for the
 localized FRBs at high $\rm DM_E$, and also accommodate the localized
 FRBs at low $\rm DM_E$ as in the fiducial model with smaller $e^\mu$,
 similar to the cases of $\ell$ in Sec.~\ref{sec3}.~Here, we still use
 the log-normal distribution of $\rm DM_{host}$
 in Eq.~(\ref{eq8}) without the location parameter $\ell$
 (namely $\ell=0$), but we instead consider a step-like $e^\mu$
 to this end, namely
 \be{eq49}
 e^\mu=
 \begin{cases}\\[-6.1mm]
 \;\mu_0 \quad & {\rm if}\hspace{2mm} {\rm DM}_{{\rm E},\,i}
 <{\rm DM_{E,\,t}}\,,\\[1.7mm]
 \;\mu_1 & {\rm if}\hspace{2mm} {\rm DM}_{{\rm E},\,i}\geq
 {\rm DM_{E,\,t}}\,.
 \end{cases}
 \ee
 We label this model as Mu2s, with three new model parameters $\mu_0$,
 $\mu_1$, ${\rm DM_{E,\,t}}$ (all in units of
 ${\rm pc\hspace{0.24em} cm^{-3}}$). One can go further with
 a three-step $e^\mu$, namely
 \be{eq54}
 e^\mu=
 \begin{cases}\\[-6.1mm]
 \;\mu_0 \quad & {\rm if}\hspace{2mm} {\rm DM}_{{\rm E},\,i}
 <{\rm DM_{E,\,t1}}\,,\\[1.2mm]
 \;\mu_1 & {\rm if}\hspace{2mm} {\rm DM_{E,\,t1}}\leq {\rm DM}_{{\rm E},\,i}
 <{\rm DM_{E,\,t2}}\,,\\[1.2mm]
 \;\mu_2 & {\rm if}\hspace{2mm} {\rm DM}_{{\rm E},\,i}\geq
 {\rm DM_{E,\,t2}}\,.
 \end{cases}
 \ee
 We label this model as Mu3s.~Note that ${\rm DM_{E,\,t2}}={\rm
 DM_{E,\,t1}}+S$ is a derived parameter, while $S\geq 0$ is a free model
 parameter.~This is just a technical trick to ensure ${\rm DM_{E,\,t2}}
 \geq {\rm DM_{E,\,t1}}$, as mentioned in Sec.~\ref{sec3}.~Additionally,
 we adopt the Taylor expansion of $e^\mu$ with respect to $\ln \rm DM_E$
 up to first order,
 \be{eq59}
 e^\mu=\mu_0+\mu_1\ln {\rm DM_E}\,.
 \ee
 We label this model as Mulin, and take the same considerations
 as for the Loclin model in Sec.~\ref{sec3}.

A varying $e^\mu$ is well motivated by simulations.~In
 e.g.~\cite{Zhang:2020mgq}, $e^\mu$ at redshifts between $0.1$ and $1.5$
 were derived from the IllustrisTNG $N$-body simulation for various
 types of FRBs, and it was found that $e^\mu$ monotonously increases
 with redshift (see Table~3 of~\cite{Zhang:2020mgq}).~This supports our
 explorations here.

Fitting the Mu2s, Mu3s, Mulin models to the data with the uniform priors
 given in Table~\ref{tabprior2}, we present the results in
 Figs.~$\ref{fig9}-\ref{fig11}$, Tables~\ref{tabcons2} and~\ref{tab3},
 respectively. From Figs.~$\ref{fig9}-\ref{fig11}$ and
 Tables~\ref{tabcons2}, we find that the derived Hubble constants $H_0$
 for these models are well consistent with the ones of Planck 2018 and
 SH0ES in the $1\sigma$ region. It is easy to see from Table~\ref{tab3}
 that all these models are overwhelmingly preferred over the fiducial
 model by the Bayesian evidence and AIC, BIC.

\begin{center}
 \begin{figure}[tb]
 \centering
 \vspace{-5.5mm}  
 \includegraphics[width=0.49\textwidth]{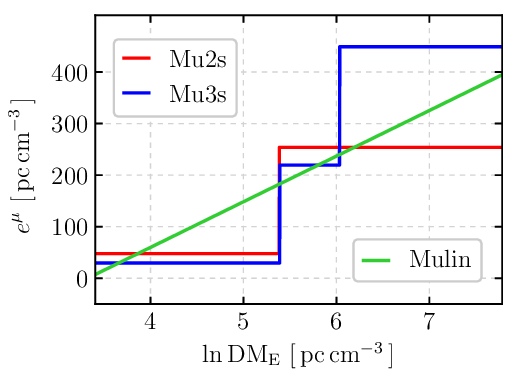}
 \vspace{-1mm}  
 \caption{\label{fig12} $e^\mu$ versus $\ln {\rm DM_E}$ for
 all the $e^\mu$ models.~See Sec.~\ref{sec4} for details.}
 \end{figure}
 \end{center}

In Fig.~\ref{fig12}, we present $e^\mu$ versus $\ln \rm DM_E$ for all
 the $e^\mu$ models, where the model parameters take their mean values
 obtained from the fits to the data.~We find that in all the step-like
 $e^\mu$ models, $\mu_0\sim 50$ or $30$ (fairly close to the $e^\mu\sim
 68$ of Macquart {\it et al.}~\cite{Macquart:2020lln} and the $e^\mu\sim
 69$ of the fiducial model in Eq.~(\ref{eq17})), and $\mu_1\sim 250$ or
 $220$, while the transition happens around $\ln {\rm DM_E}\sim
 5.4$.~The $e^\mu$ of Mulin model crosses almost all steps of $\mu_i$,
 as shown in Fig.~\ref{fig12}, and hence this linear $e^\mu$ model is
 at least a decent enough approximation of the unknown $e^\mu$
 function.~Similar to the generalized $\ell$ studies in Sec.~\ref{sec3},
 the common feature of all the $e^\mu$ models is that $e^\mu$ converges
 to a smaller value ($\sim {\cal O}(10)$) to accommodate the localized
 FRBs at low $\rm DM_E$ as in the fiducial model, and then
 $e^\mu$ becomes much larger at high $\rm DM_E$ which shifts $H_0$ to
 larger values. The physical picture is clear. In Table~\ref{tab3}, we
 also summarize the Bayesian evidences and the information criteria AIC,
 BIC for all the $e^\mu$ models relative to the fiducial model.~One can
 easily see that $\rm Mu3s\gg Mulin\gg Mu2s
 \gg Fiducial$.~\hspace{0.2mm}The Hubble constants $H_0$ can be
 consistent with the ones of Planck 2018 and SH0ES in all the $e^\mu$
 models.~Interestingly, it looks like the constraints on $H_0$ are of
 similar size when comparing the Mu2s and Mu3s models, or even better
 with the Mu3s model (we thank the referee for pointing out
 this issue).~This is somewhat counter-intuitive, since the
 latter has two additional free parameters $\mu_2$ and $S$.~The
 correlations between the model parameters $\Theta$ and $\mu_2$, $S$
 might help to break the degeneracy between $\Theta$ and the other model
 parameters, and hence we could obtain a slightly tighter constraint on
 $\Theta$.~Noting that the constraint on $H_0$ is derived from
 $\Theta$, this might explain the interesting observation mentioned
 above.


 \begin{center}
 \begin{figure}[tb]
 \centering
 \vspace{-5.4mm}  
 \includegraphics[width=0.5\textwidth]{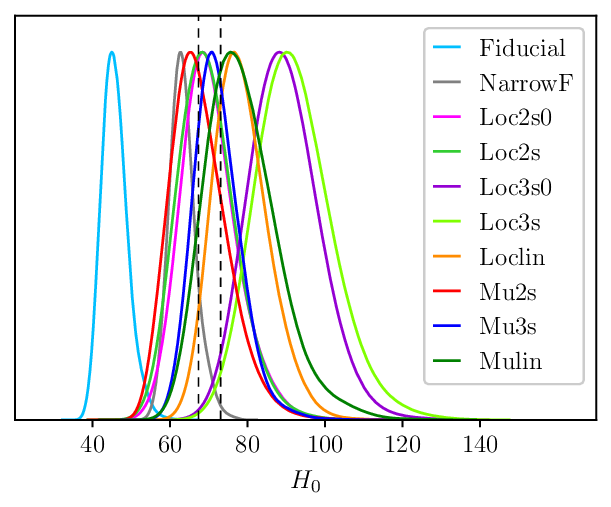}
 \vspace{-1.5mm}  
 \caption{\label{figh0all} The posteriors on $H_0$ (in units of
 $\rm km/s/Mpc$) for all models, while $H_0=67.36$ and $73.04\;
 {\rm km/s/Mpc}$ of Planck 2018 and SH0ES are indicated by
 the vertical dashed lines.}
 \end{figure}
 \end{center}


\vspace{-18mm} 


\section{Concluding remarks}\label{sec5}

In the present work, we test the robustness of the Macquart {\it et al.}
 methodology~\cite{Macquart:2020lln} (which uses FRBs as a cosmological
 probe), by allowing for more general distributions of $\rm DM_{host}$,
 while simultaneously addressing its limitation (e.g.~the parameter $F$
 in the distribution of $\rm DM_{IGM}$ is unbounded from above with the
 narrow prior) and also alleviating the Hubble tension between FRBs,
 Planck 2018 and SH0ES. In fact, a small $F<1$ is usually imposed in the
 Macquart {\it et al.} methodology, and it is the key to obtain values
 of the cosmological parameters $H_0$, $\Omega_b$ and $f_{\rm IGM}$ that
 are more consistent with the widely accepted values (note that in the
 present work $\Omega_b h^2$ and $f_{\rm IGM}$ are fixed, but they are
 free in many relevant works in the literature). In the present work, we
 consider a loose prior for the parameter $F$ by allowing $F>1$, and find an
 unusually low $H_0$ from 125 localized FRBs.~We show that the model
 with loose $F$ prior (allowing $F>1$) is strongly preferred over the
 one with narrow $F$ prior (imposing $F\leq 0.5$, at least $F<1$) using
 all three of our model comparison metrics (\hspace{0.05em}AIC, BIC and
 $\ln {\cal B}$\hspace{0.03em}), but yields a value of $H_0$ that is
 in tension with both the CMB and SNIa constraints at the $>7\sigma$
 level.~Instead of modifying $\sigma_\Delta=Fz^{-0.5}$ in the
 distribution of $\rm DM_{IGM}$ as in e.g.~\cite{Zhuge:2025urk}, here we
 explore the alternative of generalizing the distribution of
 $\rm DM_{host}$ with varying location and scale parameters $\ell$ and
 $e^\mu$, respectively.~We find that these complex models of $\ell$ and
 $e^\mu$ result in constraints on $H_0$ that are well consistent with
 the CMB and SNIa values, while simultaneously being strongly
 preferred to both the fiducial model and the original model
 used by Macquart {\it et al.}~\cite{Macquart:2020lln}.

In this work, we have considered eight models of
 the generalized distributions of $\rm DM_{host}$.~In Table~\ref{tab3},
 we summarize the Bayesian evidences and the information criteria AIC,
 BIC for all models.~We can easily see that $\rm Loc3s0>Loc3s\gg Mu3s
 \sim Loclin\gg Mulin\sim Loc2s0>Loc2s\gg Mu2s\gg Fiducial\gg NarrowF$.
 All the generalized $\rm DM_{host}$ models are overwhelmingly preferred
 over the fiducial model.~We consider that the simple Loc2s0, Mu2s,
 Mulin or Loclin models are enough in practice.~However, it is not enough to
 compare models by using only the Bayesian evidence, AIC and BIC, while
 the constraints for the different models should be also taken into
 account (we thank the referee for pointing out this issue).~In general, the
 models with more free parameters also have looser constraints on $H_0$.~But
 this is not the case in the present work.~For example, the constraints
 on $H_0$ for the Mu2s and Mu3s models are of similar size, mainly due
 to the correlations between the model parameter $\Theta$ and
 the additional free parameters $\mu_2$ and $S$ in the Mu3s model, as
 discussed at the end of Sec.~\ref{sec4}.~The Loc2s0/Loc2s and
 Loc3s0/Loc3s models are also in a similar situation, but mainly due to
 the fact that $\ell_0=0$ is still within the $1\sigma$ region
 of $\ell_0$ for the Loc2s and Loc3s models, as discussed in
 Sec.~\ref{sec3}.~Based on the constraints on $H_0$, we prefer
 the Loc2s0, Loc2s, Mu2s, Mu3s models in practice.~In addition, some of
 the corner plots show either poorly constrained posteriors
 (e.g.~$\ell_0$ for the Loclin model) or multimodal ones (e.g.~$\rm
 DM_{E,\,t}$ for the Loc2s model).~Excluding these two models,
 and considering the intersection of these two lists of the preferred
 models based on both $\ln {\cal B}$/AIC/BIC and the constraints on
 $H_0$ mentioned above, we finally recommend the Loc2s0 and Mu2s models.

In Fig.~\ref{figh0all}, the posteriors on $H_0$ for all models
 considered in this work are given.~It is easy to see that most of them
 could be well consistent with the ones of Planck 2018 and SH0ES.~Note
 that the constraints on $H_0$ from FRBs are all wider than the
 constraints using the CMB and SNIa measurements, and hence it cannot
 yet differentiate between the $H_0$ constraints, mainly due to the fact
 that only 125 localized FRBs are used in this work, while
 ${\cal O}(10^3)$ SNIa are available currently.~More localized FRBs are
 needed to obtain the competitive constraints on $H_0$, at least more
 than the number of SNIa, namely $\gtrsim {\cal O}(10^3)$. More
 precisely known probability distributions of $\rm DM_{IGM}$,
 $\rm DM_{host}$ and $\rm DM_{MW,\,halo}$ are also needed to improve the
 constraints.~The subclassification of FRBs might be helpful.~Similar to
 the field of supernovae in which only type Ia (rather than types Ib,
 Ic, IIp, IIn) supernovae could be used as standard candles for
 cosmology, it is of interest to find a suitable subclass of
 FRBs (rather than using all types of FRBs) for cosmology.~To this end,
 in e.g.~\cite{Guo:2022wpf,Guo:2023hgb,Li:2024dge}, we have proposed a
 new subclassification scheme of FRBs (different
 from repeaters/non-repeaters), in which type Ib FRBs (nyFRBs) might be
 promising.~Let us keep an open mind for the precision cosmology with
 FRBs in the future.

The parameter $F$ is related to the galactic feedback parameterized by
 $\sigma_\Delta=Fz^{-0.5}$ in the distribution of $\rm DM_{IGM}$ (see
 Eqs.~(\ref{eq6}) and (\ref{eq7})).~$F$ quantifies the strength of the
 baryon feedback, and a smaller/larger~$F$ corresponds to a
 stronger/weaker feedback~\cite{Macquart:2020lln}.~The parameterization
 $\sigma_\Delta=Fz^{-0.5}$ introduced in~\cite{Macquart:2020lln} worked
 well for their small sample of only 8 localized FRBs, while they argued
 that $F<0.5$ is reasonable enough.~But as shown by Extended Data Fig.~5
 of~\cite{Macquart:2020lln} with the prior $F\leq 0.5$, the parameter
 $F$ cannot be constrained from above, suggesting that the observational
 data prefer a larger $F>1$.~This is also borne out by our own
 findings.~Although we adopt a fairly loose prior $F\leq 10$, the means
 of $F$ for all models are in the range of $2.5\sim 4.5$ (see
 Tables~\ref{tabcons1} and \ref{tabcons2}).~$F$ can already be
 constrained with the prior $F<5\sim 6$ (see Figs.~\ref{fig1},
 $\ref{fig3}-\ref{fig7}$, $\ref{fig9}-\ref{fig11}$), and it is not needed to
 be as large as $10$.~Our results prefer a larger $F>1$ (around $1\sim 3$ is
 enough), which indicates that the galactic feedback might be weaker
 than the one commonly assumed in the literature.~One should be aware of
 this potentially serious issue (we thank the referee for pointing out
 it).~In fact, the parameterization $\sigma_\Delta=Fz^{-0.5}$ might
 not be sufficient to describe the galactic feedback, as discussed in
 e.g.~\cite{Zhuge:2025urk}.~In particular, $\sigma_\Delta\to\infty$ as
 $z\to 0$, and hence this parameterization does not work well at low
 redshifts.~In~\cite{Zhuge:2025urk}, it was argued that $\sigma_\Delta$
 should be modified and a fairly complicated form was suggested.~We consider
 that a deeper discussion on $\sigma_\Delta$ is needed but it is beyond
 the scope of the present work, since here we mainly focus on
 $\rm DM_{host}$.

In FRB cosmology, the useful combination $\Theta\propto
 (\Omega_b h^2)\,f_{\rm IGM}/H_0$ defined in Eq.~(\ref{eq4})
 characterizes the degeneracy between $H_0$, $\Omega_b h^2$ and
 $f_{\rm IGM}$.~Note that $\Omega_b h^2$ as a whole can be independently
 constrained by using e.g.~CMB or big bang nucleosynthesis (BBN).~So, in
 order to obtain a larger Hubble constant $H_0$, one should adopt a
 larger $f_{\rm IGM}$ close to its upper bound of $1$ for a
 fixed $\Theta$, or, if a Hubble constant $H_0$ from SH0ES or Planck
 2018 is adopted, one can find an unusually large
 $f_{\rm IGM}$.~Actually, this is the hidden trick in the
 literature.~So, the real key is not $f_{\rm IGM}$ or $H_0$.~The boss
 behind the curtain is $\Theta$, the pre-factor of $\langle{\rm
 DM_{IGM}}\rangle$ in Eq.~(\ref{eq4}).~One can see that $\Theta$ is
 correlated with the parameter $F$ in $\sigma_\Delta=Fz^{-0.5}$
 describing the distribution of $\rm DM_{IGM}$, as shown by the
 $\Theta-F$ contours in Figs.~$\ref{fig1}-\ref{fig7}$ and
 $\ref{fig9}-\ref{fig11}$.~In order to obtain a small $\Theta$ leading
 to a large $H_0$ with a normal $f_{\rm IGM}\sim 0.83$, a small $F$ is
 required.~So far, we can understand the hidden secret to bound the
 parameter $F$ with a small value of $0.5$ (at least $F<1$) in
 the literature.~But we consider that it is a trick more than
 a physical solution.~So, we have explored alternatives in the present work.

The physics in this work is $\rm DM_{host}$.~As shown in
 Secs.~\ref{sec3} and \ref{sec4}, the location parameter $\ell>0$ and/or
 the larger scale parameter $e^\mu$ can shift $\rm DM_{host}$ toward
 higher values,~or equivalently, shift $\rm DM_{IGM}$ toward smaller
 values for the fixed ${\rm DM}_{{\rm E},\,i}$ of a given localized FRB
 at redshift $z_i$ (n.b.~Eq.~(\ref{eq3})).~This leads to a smaller mean
 $\langle{\rm DM_{IGM}}\rangle$, so that we find a smaller $\Theta$ as
 the pre-factor of $\langle{\rm DM_{IGM}}\rangle$, and then a larger
 $H_0$ for a normal $f_{\rm IGM}\sim 0.83$ (n.b.~$\Theta\propto
 (\Omega_b h^2)\,f_{\rm IGM}/H_0$ defined in Eq.~(\ref{eq4})).~The key
 is $\Theta$ and $\rm DM_{host}$.~Everything is natural
 and reasonable in this physical picture.

Note that in the present work, we have fixed $\Omega_b h^2=0.02237$
 from the Planck 2018 result~\cite{Planck:2018vyg} and $f_{\rm
 IGM}=0.83$~\cite{Deng:2013aga,Yang:2016zbm,Gao:2014iva,Zhou:2014yta,
 Qiang:2019zrs,Qiang:2020vta,Qiang:2021bwb,Qiang:2021ljr,Guo:2022wpf,
 Guo:2023hgb} to derive the Hubble constant $H_0$ from $\Theta$
 in Eq.~(\ref{eq4}). In actuality, we are constraining the combination
 $\Theta\propto (\Omega_b h^2)\,f_{\rm IGM}/H_0$, so that one $H_0$,
 $\Omega_b h^2$ and $f_{\rm IGM}$ can all be derived from the free model
 parameter $\Theta$ when two of the three are fixed.~So, for a fixed
 $\Theta$, if $H_0$ changes (e.g.~is very different from the ones of
 Planck 2018 or SH0ES), the values of $\Omega_b h^2$ and $f_{\rm IGM}$
 derived from $\Theta$ will also be changed accordingly.~On the contrary, if
 we trust the values of $\Omega_b h^2$ and $f_{\rm IGM}$, the derived
 $H_0$ from $\Theta$ is also trustworthy.~Since $\Omega_b h^2$ can be
 independently constrained by using the observational data of CMB and
 BBN, while $f_{\rm IGM}$ can be independently constrained by
 using Ly$\alpha$ forest and UV absorption lines~\cite{Fukugita:1997bi,
 Shull:2011aa} (see also e.g.~\cite{Deng:2013aga}), adopting the fixed
 $\Omega_b h^2$ and $f_{\rm IGM}$ to derive $H_0$ from $\Theta$
 is equivalent to constraining $H_0$ by jointly using the observational
 data of FRBs, CMB, BBN, Ly$\alpha$ forest and UV absorption lines.

In our findings, the models with higher complexity are statistically
 preferred.~This begs the question of whether models with even more
 complexity (e.g.~four or more steps) would be suitable (we
 thank the referee for pointing out this issue).~In principle,
 any (unknown) analytic function (curve) could be better approximated
 using more steps, although the larger number of free parameters adds
 additionally penalties when comparing models.~However, the
 more fundamental problem is to determine more realistic or analytic
 functions for $\ell$ and $e^\mu$, using studies
 such as~\cite{Zhang:2020mgq}.

In this work, we have used the location and scale parameters $\ell$ and
 $e^\mu$ separately.~Future works could instead vary these two
 parameters simultaneously, leading to potentially interesting
 effects.~Additionally, it is of interest to find some
 reasonable (semi-)analytic functions for $\ell$ and $e^\mu$, rather
 than the simple Taylor expansions considered in the present work.~This
 might be a difficult task and deserves further study.~Another avenue of
 study is to explore, for example, generalized distributions of $\rm
 DM_{IGM}$, such as alternative forms of $\sigma_\Delta=Fz^{-0.5}$ to
 replace in the distribution of
 $\rm DM_{IGM}$~\cite{Zhuge:2025urk}.~\hspace{0.2mm}We leave
 these explorations to future works.

\vspace{-3mm} 


\section*{ACKNOWLEDGEMENTS}

We thank the anonymous referee for quite useful comments and
 suggestions, which helped us to improve this work.~We are grateful to
 Profs.~Puxun~Wu, Fa-Yin~Wang, Jun-Jie~Wei and Hai-Nan~Lin, as well as
 Dao-Hong~Gao, Jia-Lei~Niu, Shu-Yan~Long, Hui-Qiang~Liu, Wei-Zhi~Gong
 and Shuo-Yu~Zhang for kind help and useful discussions.~This work was
 supported in part by NSFC under Grants No.~12375042 and No.~11975046.

\renewcommand{\baselinestretch}{1.1}


\end{document}